\RequirePackage{fix-cm}
\documentclass{article}  %
\usepackage{amsmath,amssymb,graphicx,yfonts}

\newcommand{\be}{\begin{equation}}
\newcommand{\ee}{\end{equation}}
\newcommand{\bea}{\begin{eqnarray}}
\newcommand{\eea}{\end{eqnarray}}
\newcommand{\pr}{\partial}
\newcommand{\GZ}{\mathcal{G}_Z} 

\newcommand{\Bdet}{\mathcal{B}}
\newcommand{\Etm}{\mathfrak{E}}
\newcommand{\Btm}{\mathfrak{B}}
\newcommand{\non}{\nonumber}
\newcommand{\fwd}{\rightarrow}
\newcommand{\bwd}{\leftarrow}

\begin{document}

\title{Electromagnetic Gravitational Waves Antennas \\ for Directional Emission and Reception}

\author{Andr\'e F\"uzfa\thanks{andre.fuzfa@unamur.be}\\
Namur Institute for Complex Systems (naXys),\\ University of Namur, Belgium
}      

\maketitle

\begin{abstract}
\noindent 
A successful experiment combining emission and reception of gravitational waves (GWs) would constitute a premiere of gravity control. However, such experiments manipulating gravity would require to compactly store large amounts of energy while using ultra-sensitive detection techniques and hence can be expected to be very difficult to realise.
In this paper, we first propose new designs of electromagnetic (EM) generators allowing directional emission of GWs.
They are based on the resonant amplification of GWs occuring when EM standing waves couple to an external static magnetic field. The interplay between the EM polarizations, the orientation of the boosting magnetic field and the emitted GW polarizations is studied, as well as the gravitational radiation patterns of the different generators. Then, we develop a new GW directional detection method based on magnetic energy storage. Three possible applications are presented: (i) indirect detection of GW emission by high-accuracy monitoring of the energy loss in the generators ; (ii) direct detection of an incoming plane GW with a high-field magnet and (iii) GW emission-reception experiments. Although current technology allows reaching detection thresholds in these experiments, their practical realisation would still need to overcome several key technical challenges.
\end{abstract}

\section{Introduction}
\label{intro}
Out of the four fundamental interactions, gravitation remains the only one not to be under technological control. This puts limitations on experimental gravity: while we passively explore the permanent \textit{natural} gravitational fields generated by inertial masses, we do not attempt to \textit{artificially} bend spacetime at will in our laboratories. Actually, generating gravitational fields does not belong to science-fiction: it is a natural possibility offered by Einstein's Equivalence Principle. The universality of free fall teaches us that all types of energy, associated to any of the four fundamental forces, undergo an external gravitational field in the same way. But this also implies that all types of energies produce gravity in the same way. Since we cannot switch off the binding energies of matter (inertial) sources, one should rely on electromagnetic (EM) energies as a source of \textit{human-made} controllable gravitational fields. Controlling gravity, in an experimental sense, requires not only \textit{to generate}, but also \textit{to detect},  artificial gravitational fields. \\
\\
The problem of gravity control through the use of EM fields has been considered for decades. While  
 Weber \cite{weber} envisioned the importance of both the generation and detection of gravitational waves (GWs) as early as 1960, Gertsenshtein \cite{gertsenshtein} discovered in 1962 a wave resonance mechanism allowing to produce GWs from light waves passing through a region of static magnetic field. A decade later, this mechanism has been applied to astrophysics by Zeldovich \cite{zeldovich}. Grishchuk and Sazhin then introduced in \cite{grishchuk1,grishchuk2} purely electromagnetic generators of GWs, using transverse magnetic/electric (TM/TE) resonant cavities. 
 Their motivation was also to conduct GW emission-reception laboratory experiments and they concluded they could be feasible experimentally in \cite{grishchuk2}, considering this as the first necessary step for any future technological application of GW physics \cite{grishchuk3}. Resonant cavities and EM waveguides were then considered as possible detectors of gravitational radiation emitted 
 either by natural or artificial sources \cite{braginskii1,braginskii2,grishchuk2,grishchuk3,pegoraro,caves,gerlach}.
 
However, due to the faintness of the emission of GWs by foreseeable experimental devices, the strategy to develop GW physics was to use the much stronger astrophysical sources and to attempt detecting them in ground-based specific facilities. But this also meant to postpone the challenge of gravity control with the joint emission and reception of GWs. After decades of widespread effort, this astrophysical strategy has led to the \textit{direct measurement} (and not mere detection) of the final stage of binary black hole mergers by the LIGO and Virgo collaborations in 2015 \cite{LIGO}. 

Detection of astrophysical GWs with laser interferometers such as LIGO allows exploring rather low frequency ranges (typically less than $1-10kHz$). Electromagnetic detectors of GWs would allow exploring higher frequency range, typically from kHz to 100GHz when using radio-frequencies or from 100GHz to THz when using microwaves. 
However, there are few expected astrophysical sources above several kHz (cf. \cite{andersson}).
Yet investigation on electromagnetic detectors of GWs started in the 1970's with the works \cite{braginskii1,braginskii2,grishchuk2,boccaletti,pegoraro,caves,cruise}. Those detectors either make use of the conversion of GWs to photons \cite{boccaletti}, the excitation or modification of resonant modes of EM cavities and waveguides in \cite{braginskii1,braginskii2,grishchuk2,pegoraro,caves,ballantini}, the change of polarization plane of an EM wave due to the passing GWs \cite{cruise}, or induced birefringence of the interior of the cavity \cite{gerlach}. 

Nowadays, GWs emitted from natural sources have been directly detected and with such an accuracy that these messengers can be used to study the Universe in an unprecedented way. However, the emission and reception of GWs in laboratory or for technological applications still remains a great challenge.
Indeed, generating GWs by using oscillating electric fields in a resonant cavity in previous works \cite{grishchuk1,grishchuk2,tang} will produce an amplitude of only $10^{-44}$ for an electric field of
 amplitude $1$MV/m in a volume of $1\rm m^3$. Although we consider here other designs improving this amplitude by three orders of magnitude, this is still about twenty orders of magnitude below the amplitude of GWs from astrophysical sources that are currently detected with laser interferometers. However, GWs from astrophysical sources do arrive at a random rate and, for stellar  black hole mergers, on timescales of a few hundred milliseconds. At the opposite, the GWs emitted from EM generators provide a reliable continuous flux of GWs. Different strategies of detection can therefore be developed for the weaker electromagnetically generated GWs to make the most of their stable incoming flux.\\
\\
The pioneering works \cite{grishchuk1,grishchuk2,tang} on emission and reception of GWs by EM means  considered only resonant cavities for both the generation and the detection. Those works
finally came up considering a hollow (TM/TE) toroidal microwave cavity \cite{grishchuk2}, possibly filled with some dielectric
material \cite{tang}, but their analytical computations were limited to the immediate vicinity of the torus center where standing GWs appear. Detection of the generated GWs was also considered in \cite{grishchuk2,grishchuk3} by placing a hollow cylindrical cavity  on the symmetry axis of the generator to try catching the standing GWs through mode excitation in the cavity.
\\
\\
In the present paper, we propose new designs and methods, all based on EM technology, for both the directional emission and reception of GWs as well as three possible applications to GW physics laboratory experiments.\\
\\
First, we propose to boost the emission of GWs by the use of an external uniform magnetic field. Indeed, we will show in this paper that, by \textit{appropriately orienting} the resonant cavity or waveguide carrying an electric field $E_0$ into an external magnetic field $B_0$, specific polarizations of the emitted GW reach an amplitude of order $(4 G B_0 E_0 L^2)/(c^5\mu_0)$, which can be well above the amplitude generated by the cavity or waveguide alone, which is of order $(4 G E_0^2 L^2)/(c^6\mu_0)$ for large external magnetic field ($B_0>10^{-2}$T). We also present concepts of GW generators based either on transverse electromagnetic (TEM) waveguides or TM/TE resonant cavities. 
In addition, while the works \cite{grishchuk2,grishchuk3,tang} limited their analysis of the wave profile in the immediate vicinity of the origin of coordinates, 
we will also present numerical solutions showing the anisotropy and directional propagation of the GW emission in space for the different polarizations of the gravitational radiation, in relation with the properties of the EM generator (relative orientation of the magnetic field and cavity geometry). \\
\\
Then, we introduce a new detection method based on a magnetic energy storage device whose energy variation is of order of the GW amplitude $h$ while in \cite{grishchuk2,grishchuk3} the authors considered the energy variation of a resonant cavity which is of order $h^2$.  Finally, we consider three possible applications
of our results to (i) indirect detection of GW emission from an EM generator ; (ii) directional detection of incoming GW with magnetic energy storage and (iii) GW emission-reception experiments - the equivalent of Hertz experiment for gravity. We show how detection threshold can be reached with current technology while discussing in a non-exhaustive way some expected technical challenges.
\\
\\
The structure of this paper is as follows. In Section II, we model three new types of electromagnetic directional GW antennas, emphasizing notably the interplay between the various EM and GW polarizations at work in the process as well as illustrating the propagation of the generated GWs.
In section III, we develop a directional detection method of GWs using intense magnetic fields. 
Section IV is devoted to three experimental concepts presented as possible applications of the previous results to the development of GW physics in the lab. We finally conclude with some discussions and perspectives in section V.

\section{Directional Emission of Gravitational Waves with Electromagnetic Generators}
\label{generators}
The Gertsenshtein effect \cite{gertsenshtein} is a wave resonance mechanism in which light passing through a region of uniform magnetic field, perpendicular to the direction of light propagation, produces GWs. We will apply this effect to design GW generators whose working principle will be an EM standing wave in some resonant cavity or waveguide coupling with an external static magnetic field. \\
\\
Let us therefore work in the linearized approach of Einstein-Maxwell equations, for perturbations of a background Minkowski space:
\be
g_{\mu\nu}=\eta_{\mu\nu}+c_{\mu\nu}+w_{\mu\nu}+h_{\mu\nu}\label{decomp}
\ee
where $\eta_{\mu\nu}=\rm diag(+1,-1,-1,-1)$ is the Minkowski metric and $c_{\mu\nu},w_{\mu\nu},h_{\mu\nu}\ll 1$ represent the metric perturbations respectively due to (i) the external static magnetic field generated by some coil ($c_{\mu\nu}$), (ii) the EM wave ($w_{\mu\nu}$) and (iii) the coupling between the external magnetic field and the EM wave ($h_{\mu\nu}$). Case (i) has been studied in \cite{fuzfa} but does not give rise to GW since the outer magnetic field is considered static. Case (ii) has been studied for light propagating pulses in \cite{tolman,ratzel} and in \cite{grishchuk1,grishchuk2,tang} for EM waves in resonant hollow cavities \footnote{In \cite{grishchuk1}, a special case of hollow spherical cavity in an outer radial magnetic field is briefly considered, giving rise to an admixture of terms $w_{\mu\nu}$ (case (ii)) and $h_{\mu\nu}$ (resonance - case (iii)) which were not identified as such nor exploited by the authors.}. Case (iii) actually corresponds to the resonance between the EM waves and the outer magnetic field applied to light in \cite{gertsenshtein,zeldovich} and will keep our interest here because of the amplification of the outcoming GWs it provides.  

Assuming Lorenz gauge condition $\partial_\mu h^{\mu\nu}=0$, the linearized Einstein equations can be written down (in S.I. units):
\be
\Box^2 h_{\mu\nu}=-\frac{16\pi G}{c^4} T^{\left(\rm res\right)}_{\mu\nu}  \label{direct_GZ}
\ee
where $T^{\left(\rm res\right)}_{\mu\nu}$ is the stress-energy tensor for the resonance, which is given by
\be
T^{\left(\rm res\right)}_{\mu\nu}=T^{\left(\rm tot\right)}_{\mu\nu}-T^{\left(\rm c \right)}_{\mu\nu}-T^{\left(\rm w\right)}_{\mu\nu}\label{direct_GZ2}
\ee
where $T^{\left(\rm X \right)}_{\mu\nu}$ ($X=\rm c, w, tot$) are the Maxwell stress-energy tensors in Minkowski space for the EM fields of the coil, the EM wave and their superposition, respectively:
\bea
T^{\left(\rm X\right)}_{\mu\nu}&=&-\frac{1}{\mu_0}\left(\eta^{\alpha\beta}F_{\mu\alpha}^{\rm (X)}F_{\nu\beta}^{\rm (X)}-\frac{1}{4}\eta_{\mu\nu}F_{\alpha\beta}^{\rm (X)}F^{\alpha\beta}_{\rm (X)}\right)\label{T0}
\eea
with the total (antisymmetric) Faraday tensor $F_{\mu\nu}^{\rm(tot)}=F_{\mu\nu}^{\rm(c)}+F_{\mu\nu}^{\rm(w)}$ being due to the superposition of the EM fields of the coil and of the wave. $T^{\left(\rm res\right)}_{\mu\nu}$ is traceless (from the basic properties of Maxwell stress-energy tensor) and so does $h_{\mu\nu}$ through Eq.(\ref{direct_GZ}) in Lorenz gauge only. In this gauge, the algebraic structure of the metric perturbation tensor $h_{\mu\nu}$ is identical to the one of the stress-energy tensor  $T_{\mu\nu}$ since Einstein equations Eq.(\ref{direct_GZ}) in Lorenz gauge are decoupled. This will allow to directly relate the excited polarizations of GWs (in the given coordinates) to the polarizations present in the EM wave acting as a source. Furthermore, Lorenz gauge condition is compatible with Eq.(\ref{direct_GZ}) provided the source $T^{\rm (res)}_{\mu\nu}$ verifies 
$\partial_\mu T^{\mu\nu}_{\rm (res)}=0$ or, equivalently, that the zeroth order components $F_{\mu\nu}^{\rm X}$ of the Faraday tensor are solutions of the Maxwell equations in the vacuum of Minkowski space:
 \be
 \partial_\mu F^{\mu\nu}_{\rm (\rm c,w)}=0\cdot\label{maxwell}
 \ee 
 
Equations (\ref{direct_GZ}-\ref{maxwell}) describe the direct Gertsenshtein effect: the conversion of an EM wave into a GW in the presence of an external magnetic field. We can now apply these equations to the modeling of GWs generators where EM standing waves of different polarizations will couple to an external magnetic field to produce specific metric perturbations.

The EM standing waves ($F^{\mu\nu}_{\rm (\rm w)}$) will be produced by excitation of a waveguide or a cavity whose axis of symmetry is considered lying along the $z$ direction. Choosing the length of the waveguide/cavity $L_z$ as the characteristic scale of the problem, the solution of linearized Einstein equations in Lorenz gauge Eq.(\ref{direct_GZ}) is given by the retarded potentials
\be
h_{\mu\nu} (\vec{R},T)=-\GZ \int_{\rm gen} \frac{f_{\mu\nu}(\vec{R}',T-||\vec{R}-\vec{R}'||)}{||\vec{R}-\vec{R}'||}d^3R'\label{sol_h}
\ee
where $\vec{R}$ is the dimensionless spatial position vector: $(\vec{R},T)=(X,Y,Z,T)=(x,y,z,c\cdot t)/L_z$ and where $\GZ$ is a dimensionless constant we will introduce below. In Eq.(\ref{sol_h}), the triple integral is performed over the generator, which is the volume of the waveguide/cavity entirely immersed into the external magnetic field ($F^{\mu\nu}_{\rm (\rm w)}$ vanishes outside of the waveguide/cavity). In Eq.(\ref{sol_h}), $f_{\mu\nu}=\left(c\mu_0\right)/\left(B_0 E_0\right)T_{\mu\nu}^{\rm res}$ is the dimensionless stress-energy tensor with $E_0, \; B_0$ standing for the amplitude of the electric component of the standing wave and the external magnetic field respectively. 

We dub the dimensionless number $\GZ$ in Eq.(\ref{sol_h}),
\be
\GZ=\frac{4 G B_0 E_0 L_z^2}{c^5\mu_0}\label{GZ} 
\ee
the \textit{Gertsenshtein-Zeldovich number} (GZ) associated to a given GWs generator (see also \cite{gertsenshtein,zeldovich} for dimensional considerations on the direct Gertsenshtein effect). This GZ number measures the efficiency of the generator as the instantaneous rate of conversion of EM waves into gravitational ones. The amplitude of the generated metric perturbations $h_{\mu\nu}$ are roughly of order of magnitude $\GZ$, up to some geometrical factor related to the shape of the generator. 

If we take $L_z=1\rm m$, $B_0=10\rm T$ and $E_0=1\rm MV/m$ we find $\GZ\sim 10^{-39}$. As a matter of comparison, the EM standing GWs generator in \cite{grishchuk2} only consists of a simple hollow toroidal TM or TE cavity, such that the amplitude of the generated GWs can be obtained from Eq.(\ref{GZ}) with $B_0=E_0/c$. For an electric field of amplitude $E_0\sim 1\rm MV/m$ and a characteristic length $L_z$ of one meter, we find that $h_{\mu\nu}\sim 10^{-42}$. Therefore, in a GW generator relying solely on toroidal EM waves, an electric field of about $10^{10}$V/m is required to beat the efficiency of the proposed model of generator with an external magnetic field of $10$T.\\
\\
One could turn to extreme power laser pulses, with  electric fields as high as $10^{15}$V/m over a surface of $1\rm cm^2$, to generate gravitational perturbations. However, EM pulses alone do not excite the transverse GW modes but only longitudinal ones (see also \cite{tolman,ratzel} and Eq.(\ref{h_TEM}) below). Another possible GW generator could be to shoot extremely powerful laser pulses into a transverse (pulsed) magnetic field. However, since the size of the wave packet is only a few wavelengths long, $\GZ\sim 10^{-41}$ for $E_0=10^{15}\rm V/m$, $B_0=100\rm T$ and $L_z\sim 10^{-6}\rm m$. Therefore, using a waveguide or a resonant cavity inside a DC high-field magnet seems to us the best choice
for the electromagnetic generation of GWs, with higher efficiency $\GZ$ but also offers the possibility to have a continuous emission for long durations.

We can now particularize this discussion to different types of EM standing waves in waveguides and hollow cavity and find the associated gravitational radiation polarizations and propagation.

\subsection{TEM wave resonance}

We will work in cartesian harmonic coordinates $(x^0,x^1,x^2,x^3)=(ct,x,y,z)$ and consider a TEM standing wave arising from the superposition of two progressive waves counter-propagating along the z-direction:
\bea
c.F^{(\rm w), (0)}_{01}&=& E_x^\bwd+E_x^\fwd \non\\
c.F^{(\rm w), (0)}_{02} &=& E_y^\bwd+E_y^\fwd  \non \\
c.F^{(\rm w), (0)}_{13} &=& -E_x^\bwd+E_x^\fwd\non \\
c.F^{(\rm w), (0)}_{23} &=&  -E_y^\bwd+E_y^\fwd
\eea
This TEM standing wave is assumed here immersed into a purely transverse static magnetic field ($B_z=0$):
\bea
F^{(\rm c), (0)}_{13}&=& B_y \non\\
F^{(\rm c), (0)}_{23} &=& -B_x\cdot  \non 
\eea
This gives rise to the following source of the direct Gertsenshtein effect Eq.(\ref{direct_GZ})
\bea
\tau_{00}&=& -B_x \left(E_y^\bwd-E_y^\fwd\right)+B_y \left(E_x^\bwd-E_x^\fwd\right)=\tau_{00}^{\bwd}+\tau_{00}^{\fwd}\non\\
\tau_{11}&=& B_x \left(E_y^\bwd-E_y^\fwd\right)+B_y \left(E_x^\bwd-E_x^\fwd\right)=-\tau_{22}\non\\
\tau_{12}&=& -B_x \left(E_x^\bwd-E_x^\fwd\right)+B_y \left(E_y^\bwd-E_y^\fwd\right)=\tau_{21}\label{Tmunu}
\eea
where $\tau_{33}=\tau_{00}$ and $\tau_{03}=\tau_{00}^{\bwd}-\tau_{00}^{\fwd}$ ($\tau_{\mu\nu}=\mu_0 T_{\mu\nu}^{\rm (res)}/c$).
Given the above structure of $T_{\mu\nu}^{\rm (res)}$, the resulting metric perturbations must have the form\footnote{In the vacuum outside of the GW generator, it is possible to perform a coordinate change toward the TT gauge so that one retrieves the usual transverse GW polarizations $h_{+,\times}$ as a function of the perturbations considered here (see also \cite{hobson} for instance). In fact, the emitted gravitational radiation can be expressed far from the source as a spherical transverse wave propagating radially, in the same way the radiation emitted by an electromagnetic antenna takes the form of a TEM wave propagating radially in the far field regime (see also \cite{balanis}).}:
\be
h_{\mu\nu}^{\rm TEM}=\left(
\begin{matrix}
h_\bwd+h_\fwd & 0 & 0 & h_\bwd-h_\fwd\\
0 & h_{xx} & h_{xy} & 0\\
0 & h_{xy} & -h_{xx} & 0\\
h_\bwd-h_\fwd & 0 & 0 & h_\bwd+h_\fwd
\end{matrix}
\right)\cdot
\label{h_TEM}
\ee

We can now apply the above developments to two specific TEM waveguides: (i) two coaxial cylinders and (ii) two embedded conducting boxes, both in an external transverse magnetic field. We assume the edges of the waveguides and cavities to be perfect conductors all along this section.
We also consider the TEM waveguides are aligned along the z-axis, of length $L_z$, and located between $z=-L_z/2$ and $z=+L_z/2$.

Following \cite{jackson,griffiths}, we have the following field configurations for a pure monochromatic mode of the EM fields 
\be
\left(\vec{E},\vec{B}\right)(x,y,z,t)=\left(\vec{E},\vec{B}\right)_\perp(x,y) \exp\left(\pm i(kz-\omega t)\right)\label{decomp_EB}
\ee
with $(\vec{E},\vec{B})_\perp(x,y)$ stands for the transverse components of the electric (magnetic) field, $k$ the wavenumber,
$\omega=2\pi \nu$ the pulsation of the wave with frequency $\nu$ and a $+$ ($-$) sign in the exponential indicates a wave propagating forward (backward) the z-axis. For TEM waves, 
the dispersion relation is as in vacuum, $k=\frac{\omega}{c}$, and the transverse magnetic field inside the waveguide is given by $\vec{B}_{\rm TEM}=\pm\frac{1}{c}\vec{e}_z\wedge \vec{E}_{\rm TEM}\cdot$ In this case, Maxwell equations reduce to the two dimensional Laplace equations: $\nabla_{(x,y)}^2\vec{E}_\perp=\nabla_{(x,y)}^2\vec{B}_\perp=\vec{0}$ so that both $\vec{E}_\perp$ and $\vec{B}_\perp$ derive from a scalar potential $\psi(x,y)$ also satisfying 2D Laplace equation ($\nabla_{(x,y)}^2\psi=0$). The boundary conditions for perfectly conducting surfaces, $\vec{E}_{\parallel}=\vec{B}_\perp=\vec{0}$, are fulfilled in the TEM mode when $\psi=\rm cst$ in both the inner and outer conductors.  

\begin{figure}[ht!]
\includegraphics[trim={10cm 4cm 5cm 4cm},clip,scale=0.2]{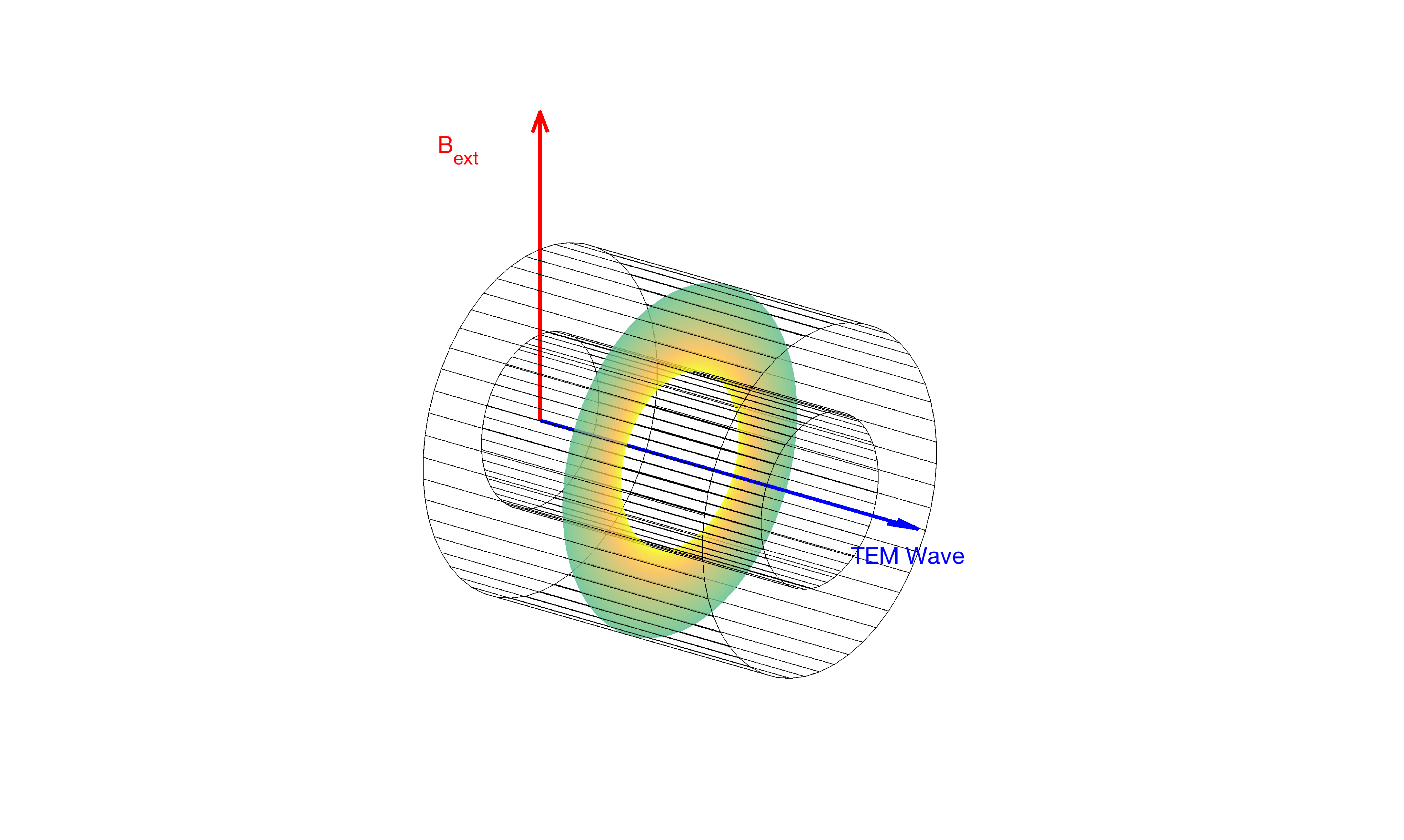}
\caption{A schematic view of GW generator: a coaxial TEM waveguide, made of two embedded
conducting cylinders, immersed into an external static magnetic field perpendicular to the axis of TEM wave propagation. The transverse slice represents the EM energy density inside the TEM waveguide, which is maximal at the interior conductor}
\label{fig_gen_cyl}
\end{figure}
\begin{figure*}[ht!]
\includegraphics[trim={6cm 11cm 6cm 2.5cm},clip,scale=0.22]{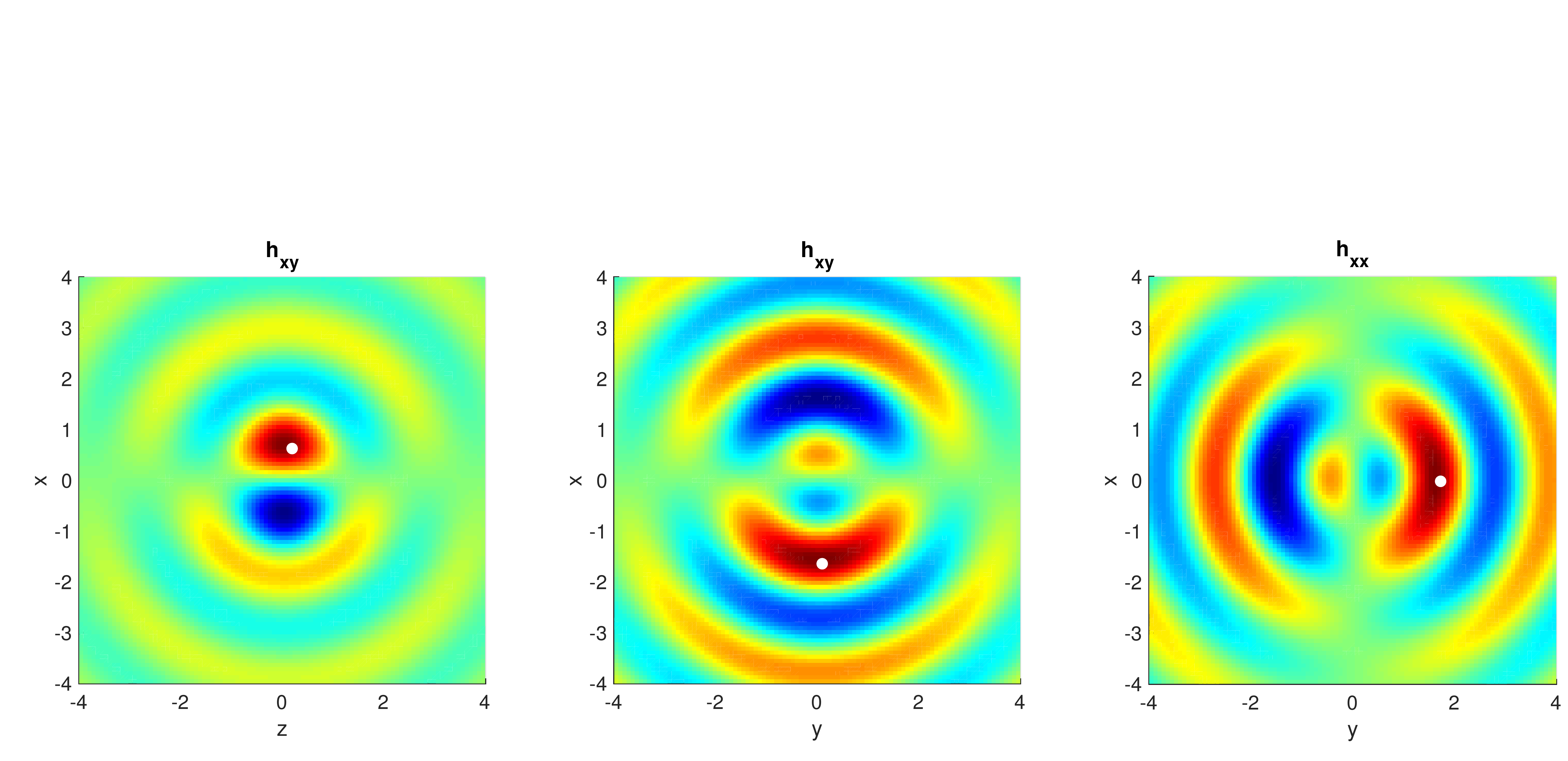}
\caption{Longitudinal ($y=0$) and transverse ($z=1$) sections of the perturbations $h_{xy}$ and $h_{xx}$
for the generator with the cylindrical TEM waveguide at some given time (with $R_{\rm in}=0.2$, $R_{\rm out}=1$ and $\lambda_{\rm GW}=2L_z$). A circle indicates the position
of the maximum of the wave pattern}
\label{sec_h12_cyl}
\end{figure*}

\subsubsection{Coaxial TEM waveguide with circular cross-section}
For the simple case of a TEM waveguide made of two embedded cylinders of inner (outer) radius $R_{\rm in}$ ($R_{\rm out}$), the solution of Maxwell equations with the above boundary conditions for the standing TEM waves is given by, in cylindrical coordinates $(r,\theta,z)$ (see also \cite{griffiths}):
\bea
E_r&=&E_0 \frac{R_{\rm in}}{r} \cos(K.ct).\sin(K.z) \nonumber\\
B_{\theta}&=&-\frac{E_0}{c} \frac{R_{\rm in}}{r}  \sin(K.ct) \cos(K.z)\label{TEM_cyl}
\eea
where $E_0$ is the electric field amplitude on the inner conductor at $r=R_{\rm in}$ and $K=2\pi/\lambda$
is the wavenumber of the TEM standing wave. The above standing waves are formed from the superposition of two sinusoidal pure modes counter-propagating in the $z$ direction. 

We can now consider the direction of the external static magnetic field as our $x$-axis so that $B_x=B_0$, $B_y=0$ and the source of GWs Eqs.(\ref{Tmunu}) now becomes
\bea
\tau_{00}&=& -B_0\left(E_y^\bwd- E_y^\fwd\right)=-\tau_{11}=\tau_{22}\non\\
\tau_{12}&=& -B_0 \left(E_x^\bwd-E_x^\fwd\right)=\tau_{21}\label{Tmunu_cyl}
\eea
and $E_{x,y}$ are obtained from Eq.(\ref{TEM_cyl}). 
This generator therefore produces GWs with polarizations $h_\bwd+h_\fwd=-h_{xx}\cdot$
A schematic view of this GW generator is given in Fig.\ref{fig_gen_cyl}.

We now present in more details the gravitational field perturbations generated by such a TEM waveguide inside a DC high-field magnet. Fig. \ref{sec_h12_cyl} presents longitudinal and transverse slices in the field distribution $h_{\times}$ 
(in units of $\GZ$) at some given time around a generator with $R_{\rm in}=0.2$, $R_{\rm out}=1$ and $\lambda_{\rm GW}=2L_z$ the outgoing gravitational radiation wavelength. Spatial dimensions are given in units of $L_z$, the length of the waveguide.
The GW emission appears anisotropic and a privileged direction is given by the direction of
the external magnetic field $\vec{B}$: the extrema of $h_{xy}=h_{12}$ points in the direction of the external field $\vec{B}$ while those of
$h_{xx}=h_{11}$ lie in direction perpendicular to $\vec{B}$ (the $x$-axis by assumption here). It should be also noticed that the cone of emission is rather large around the direction of the external magnetic field
$\vec{B}$.

\begin{figure}[ht!]
\includegraphics[trim={5cm 0cm 0 0.5cm},clip,scale=0.25]{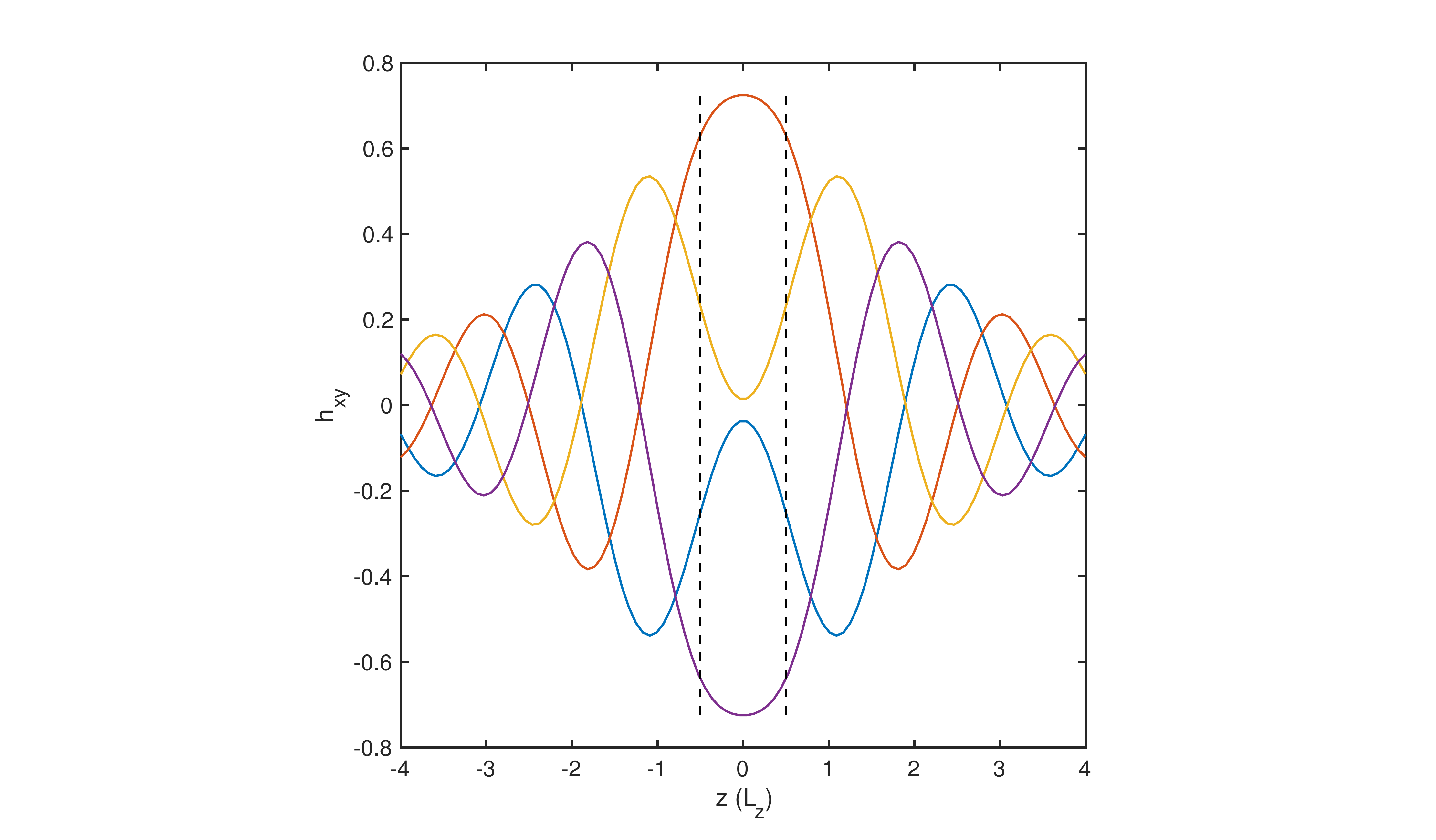}
\caption{Metric perturbations $h_{12}=h_{xy}$ at different times along the longitudinal direction $z$ for $x=1.5$ and $y=0$
for a cylindrical TEM cavity with $R_{\rm in}=0.5$ and $R_{\rm out}=1$ and $\lambda_{\rm GW}=2L_z$. Vertical lines illustrate the position of the generator along $z$}
\label{fig2}
\end{figure}

Fig. \ref{fig2} illustrates the propagation of metric perturbations $h_{xy}=h_{12}$ along the direction of the generator at different times in the interval of the wave period. The wavelength and frequency of the continuously emitted GWs are the same as the one of the EM standing wave inside the generator. This is due to the resonance mechanism, since the coupling between the variable EM field and the external magnetic field is linear (see the source Eq.(\ref{Tmunu})). The GWs $w_{\mu\nu}$ which are due to EM standing wave alone (see Eq.(\ref{decomp})) have a period which is double the one of the EM wave, because their source term Eq.(\ref{T0}) is quadratic in the EM field.

While the central wave pattern immediately in front of the generator is that of a stationary wave (as already found in \cite{grishchuk2}), the outgoing GWs quickly fade away with inverse power of the distance as the progressive waves propagate. 

However, due to rotational continuous symmetry of the waveguide around its axis, it is not possible to completely disentangle all the emitted polarizations $h_{xx,xy,\bwd,\fwd}$, as can be seen from Eq.(\ref{Tmunu_cyl}). Since choosing both the direction of emission and the emitted polarizations is important for detection and other practical applications (see section III), we can propose the following modified design.

\subsubsection{Coaxial TEM waveguide with rectangular cross-section}

\begin{figure}[ht!]
\includegraphics[trim={2cm 3cm 2cm 2cm},clip,scale=0.4]{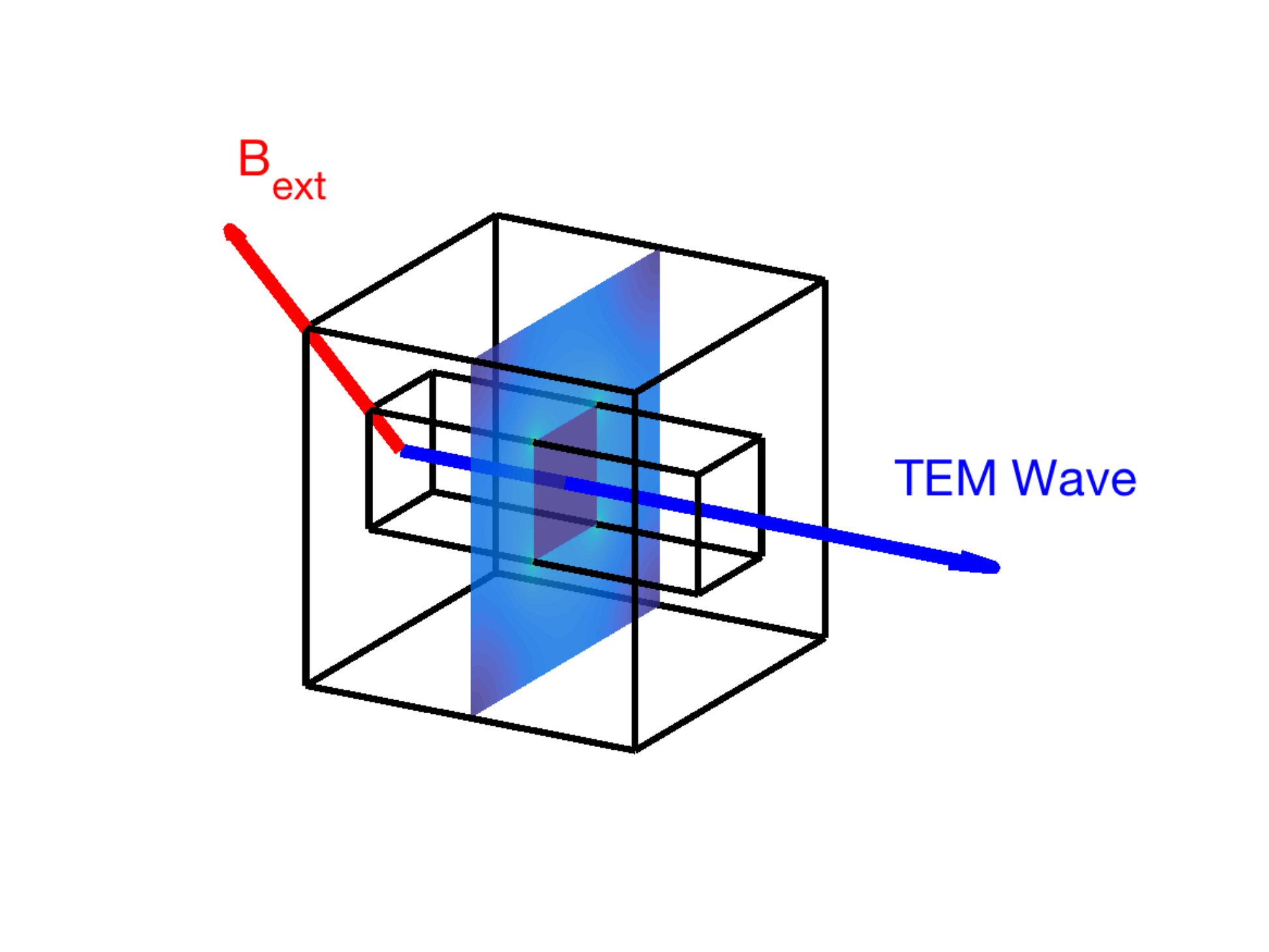}
\caption{A schematic view of the electromagnetic GW generator: a coaxial TEM cavity, made of two embedded
conducting boxes, immersed into an external static magnetic field perpendicular to the axis of TEM wave propagation. A transverse section of the EM energy density of the waveguide is also represented}
\label{fig_gen_rect}
\end{figure}

We now consider a TEM waveguide with rectangular transverse sections aligned along the $x$ and $y$-axis and of outer (inner) length(es) $L_{x,y}$ ($l_{x,y}$). The transverse external magnetic field $\vec{B}$ now has two independent components $B_{x,y}$.
This simple set-up is schematically presented in Fig.\ref{fig_gen_rect} and allows more freedom in the emitted polarizations, according to Eqs.(\ref{Tmunu}).
Therefore, using such a rectangular cross-section has two advantages: (i) it will become possible to excite the different GW polarizations more independently and (ii) directing the GW emission can be made by simply changing the relative orientation of the external magnetic field and the waveguide. In the previous design with a circular cross-section, it was necessary to rotate the outer DC magnet if one wants to change the direction of emission, which can be quite inconvenient in practice due to the size and weight of such electromagnets.
A schematic view of the generator made of a TEM rectangular waveguide coupled to an external magnetic field is given in Fig.\ref{fig_gen_rect}.
\begin{figure}[ht!]
\includegraphics[trim={2cm 0cm 0 0.5cm},clip,scale=0.2]{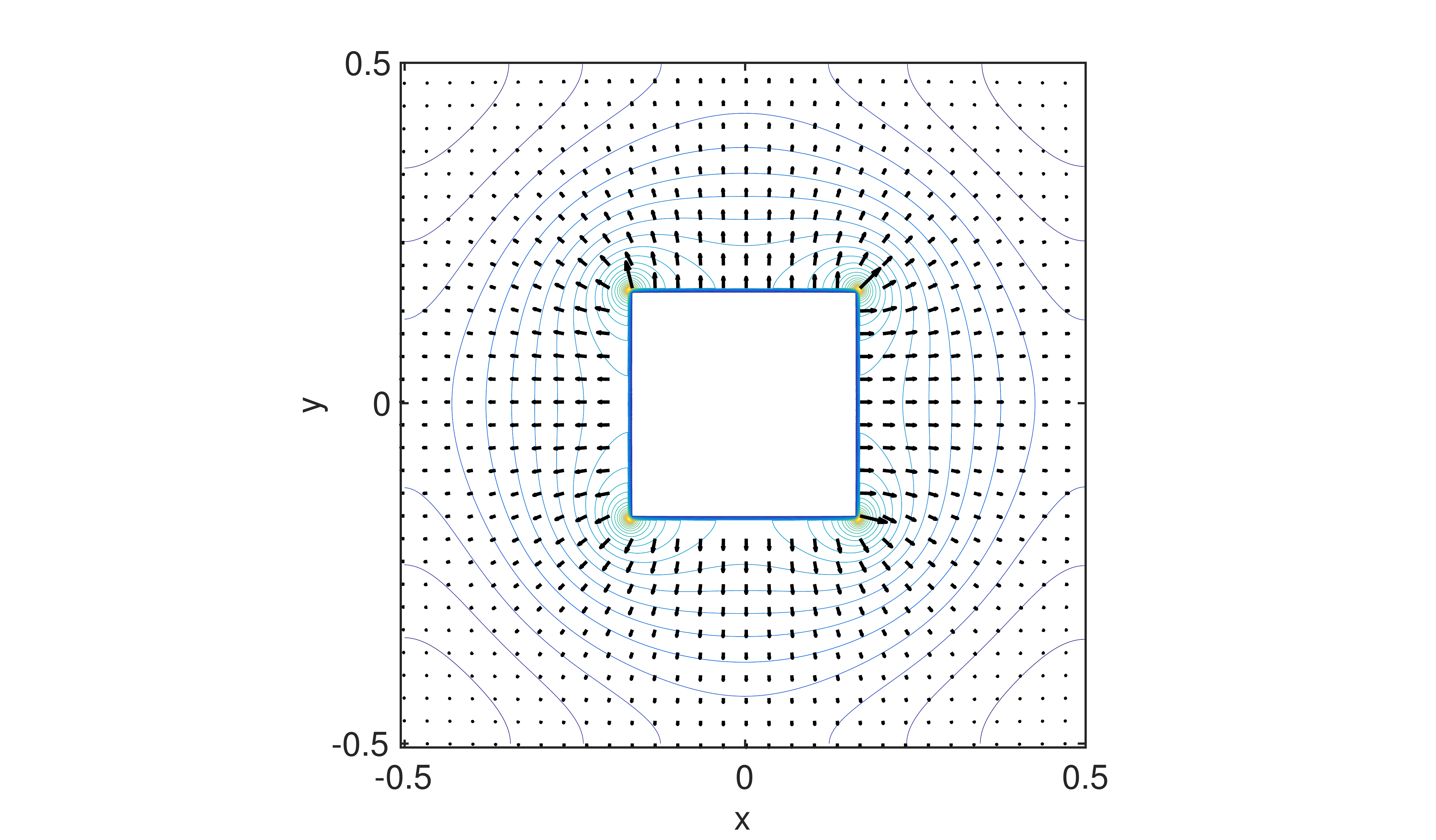}
\caption{Transverse electric field $\vec{E}_\perp$ (arrows) and the contour levels of its norm in the transverse section of the rectangular coaxial waveguide  with $L_x/l_x=L_y/l_y=3$. Edge effects around the inner conductor can be clearly viewed }
\label{fig_Et}
\end{figure}

Fig. \ref{fig_Et} shows the distribution of the transverse electric field $\vec{E}_\perp$ from a numerical resolution of the Laplace equation. This distribution can be put into Eqs.(\ref{sol_h})  and (\ref{Tmunu}) together with
the assumption of a static uniform transverse magnetic field $\vec{B}$ to find the metric perturbations around this generator.

\begin{figure}[ht!]
\includegraphics[trim={2cm 2cm 0cm 1cm},clip,scale=0.4]{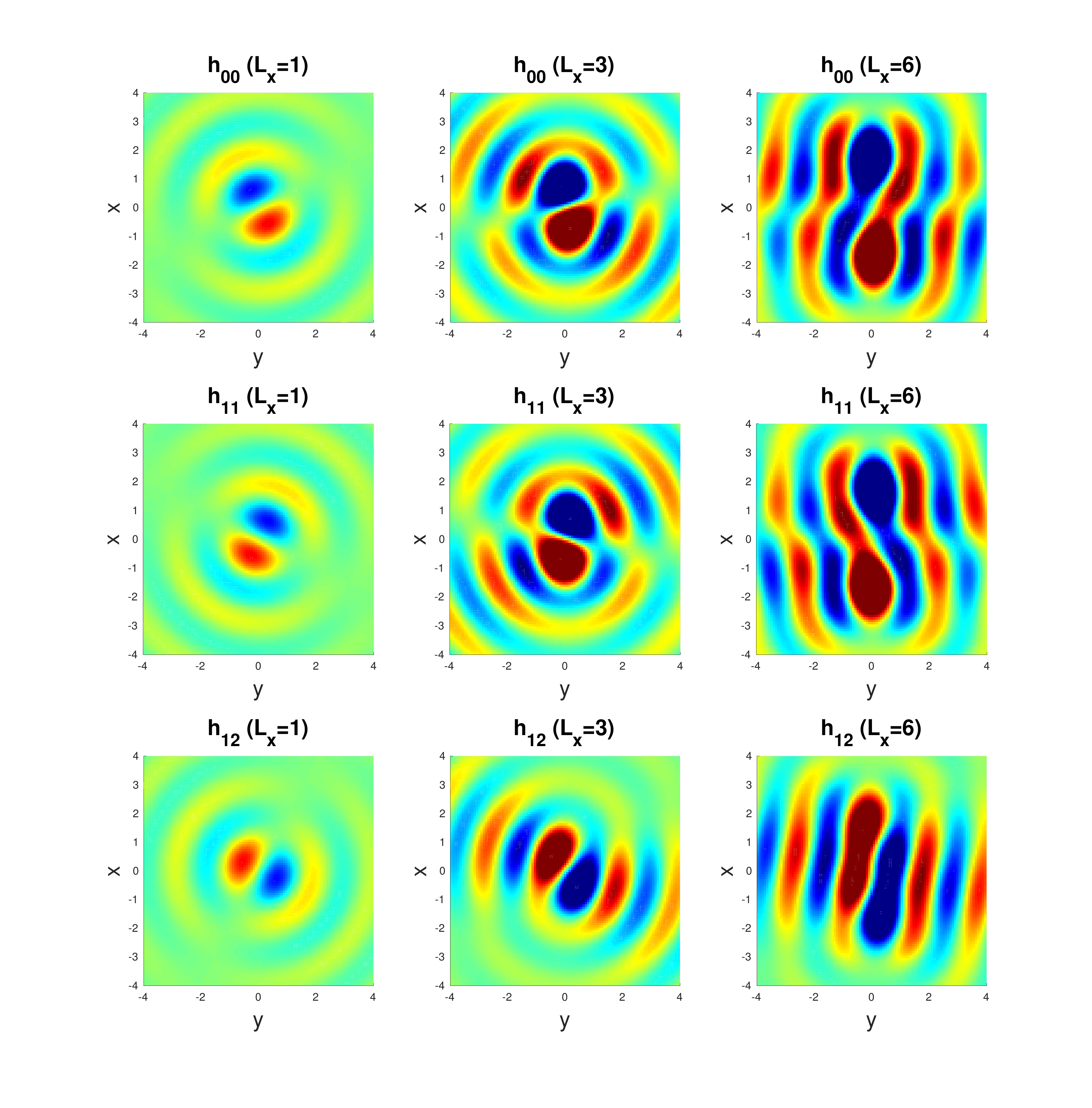}
\caption{Transverse sections at $z=0$ of the emitted GWs $h_{00}$ (first row), $h_{11}$ (middle row) and $h_{12}$ (bottom row) 
around a TEM waveguide with rectangular cross-section with various dimensions $L_x=1$ (left column)
$L_x=3$ (center) and $L_x=6$ (right column) ($L_y=L_z=1$, $L_x/l_x=L_y/l_y=3$) 
and a transverse magnetic field not aligned with the waveguide axis ($B_x\approx 0.45$, $B_y\approx 0.89$)} 
\label{sec_hs_rect}
\end{figure}

\begin{figure}[ht!]
\includegraphics[trim={2cm 2cm 0cm 1cm},clip,scale=0.4]{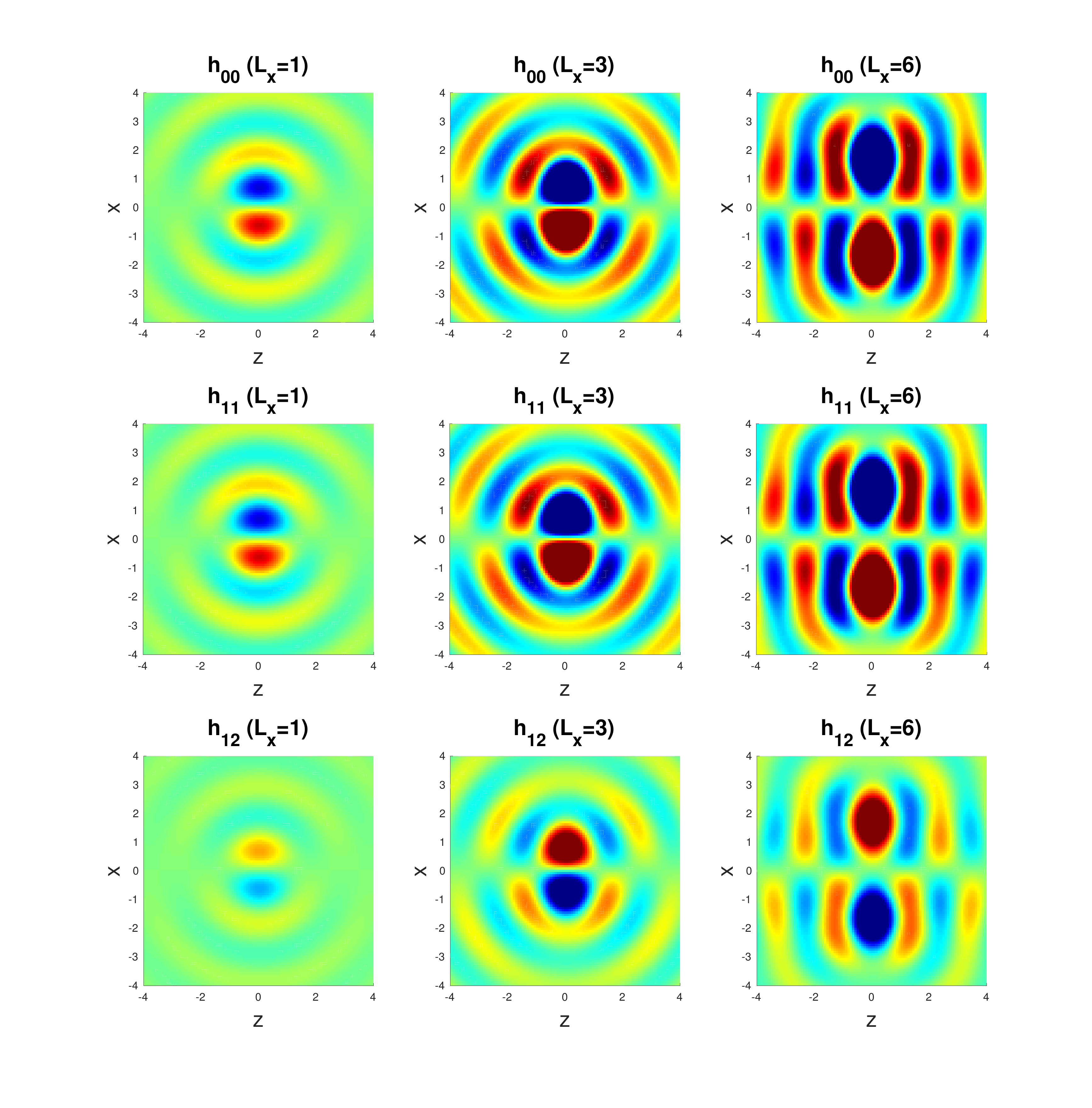}
\caption{Longitudinal sections at $y=0$ of the emitted GWs $h_{00}$ (first row), $h_{11}$ (middle row) and $h_{12}$ (bottom row) 
around a TEM waveguide with rectangular cross-section with various dimensions $L_x=1$ (left column)
$L_x=3$ (center) and $L_x=6$ (right column) ($L_y=L_z=1$, $L_x/l_x=L_y/l_y=3$) 
and a transverse magnetic field not aligned with the waveguide axis ($B_x=0.45$, $B_y=0.89$)} 
\label{sec_hs_rect_long}
\end{figure}

Figs. \ref{sec_hs_rect} and \ref{sec_hs_rect_long} present  transverse ($z=0$) and longitudinal ($y=0$) sections in the field distributions $h_{00},h_{11}$ and $h_{12}$ 
(in units of $\GZ$) at some given time around a waveguide with growing transverse length $L_x=1,3,6$ ($L_y=L_z=1$ and $L_x/l_x=L_y/l_y=3$) immersed into some arbitrary orientation of the external magnetic field $\vec{B}$. For $L_x=1$, the wave pattern looks like the one generated with a cylindrical waveguide (see Fig.\ref{sec_h12_cyl}), except that the GW polarizations are now excited differently. The direction of the GW emission now depends on the relative orientation of the external magnetic field and the waveguide transverse axis. 
It is therefore possible to direct the emission of the different GW polarizations by simply rotating the waveguide inside the external magnetic field during the functioning of the generator.   

In addition, increasing the size of the generator not only modifies the shape of the wave patterns (see Fig. \ref{sec_hs_rect}) but also increases the amplitude of the emitted GWs, since the energy stored in the waveguide is more important (the integration in Eq.(\ref{sol_h}) is then performed over a larger volume). For instance in Figs.\ref{sec_hs_rect} and \ref{sec_hs_rect_long}, the amplitude of metric perturbations has increased by roughly a factor $5$ when the transverse length of the waveguide $L_x$ has been increased from $1$ to $6$. 
\subsection{TM wave resonance}
Let us consider a generator made of a hollow resonant cavity immersed into an external static magnetic field $\vec{B}$. The cavity is assumed filled with TM modes\footnote{TE modes do not sensibly change the results presented here.}, with EM fields $F_{0i}=\Etm_i/c$, $F_{ij}=-\epsilon_{ijk}\Btm_k$ ($i,j,k=1,2,3$ and 
$\epsilon_{ijk}$ the Levi-Civita symbol) and $\Btm_z=0$. 

The source terms for the wave resonance Eq.(\ref{direct_GZ2}) are given by
\bea
\tau_{00}&=&-\left(B_x \Btm_x+B_y \Btm_y\right)=\tau_{33} \non\\
\tau_{10}&=&B_z \Etm_y-B_y \Etm_z\non\\
\tau_{20}&=&-B_z \Etm_x+B_x \Etm_z\non\\
\tau_{30}&=& B_y \Etm_x-B_x \Etm_y\non\\
\tau_{11}&=&B_x \Btm_x - B_y \Btm_y =-\tau_{22}\label{TmunuTM}\\
\tau_{12}&=&B_y \Btm_x + B_x \Btm_y \non\\
\tau_{13}&=& B_z \Btm_x\non\\
\tau_{23}&=& B_z \Btm_y\non
\eea
In the Lorenz gauge, the linearized Einstein equations decouple so that the metric perturbation tensor $h_{\mu\nu}$ must have the same algebraic structure as the source tensor Eq.(\ref{TmunuTM}). Consequently, 8 different GW polarizations, including along the axis of the device, can be excited in the chosen gauge with this generator design.

\begin{figure}[ht!]
\includegraphics[trim={10cm 4cm 5cm 4cm},clip,scale=0.15]{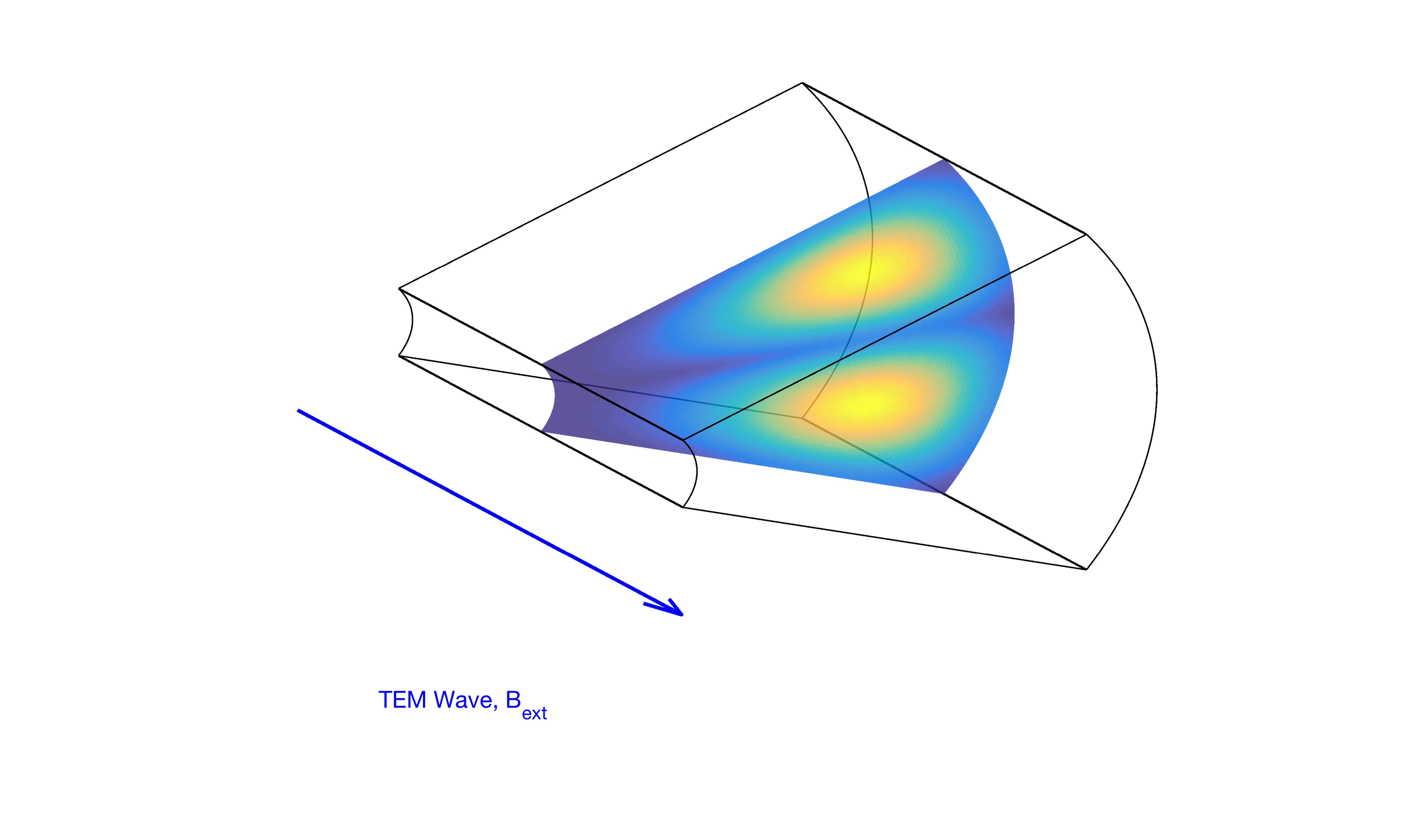}
\caption{A schematic view of a GW generator with TM wave resonance. The hollow TM cavity is made of two connected portions of cylinders while the external static magnetic field $\vec{B}$ is aligned along the axis of TM wave propagation. The transverse slice represents the EM energy density inside the cavity
for a specific TM mode}
\label{fig_gen_opencyl}
\end{figure}

Let us now consider a special case with a hollow cavity made of two connected open cylinders, namely the interior of the cavity is the region defined in cylindrical coordinates by $r\in[R_{\rm in},R_{\rm out}]$, $\theta\in[-\Theta/2,\Theta/2]$ and $z=\in[-L_z/2,L_z/2]$. For the sake of simplicity, we will restrict ourselves to an external magnetic field oriented along the longitudinal axis of the cavity: $\vec{B}^T=(0,0,B_z)$.
A schematic view of this GW generator using TM wave resonance is given in Fig.\ref{fig_gen_opencyl} for $R_{\rm in}=0.2$, $R_{\rm out}=1$, $\Theta=\pi /3$.

For hollow cavities with a uniform cross-section along their longitudinal axis $z$, TM/TE monochromatic modes are decomposed along transverse and longitudinal components as in Eq.(\ref{decomp_EB}) and the transverse components $(\vec{\Etm},\vec{\Btm})_\perp$ must satisfy the following two-dimensional Helmholtz equation
(see \cite{jackson,griffiths})
$$
\left(\nabla^2_\perp+(\omega^2-k^2)\right)\left\{\vec{\Etm},\vec{\Btm}\right\}_\perp=0
$$
with boundary conditions for perfectly conducting surfaces: $\vec{\Etm}_{\parallel}=\vec{\Btm}_\perp=\vec{0}$.
The EM fields are therefore eigenfunctions of the two-dimensional laplacian and
the dispersion relation  $\omega=\omega(k)$ can be obtained from the resolution of the related eigenvalue problem. Pure monochromatic TM standing waves in a hollow cavity made of two connected open cylinders can be obtained from the longitudinal component $\Etm_z$:
\bea
\Etm_z(r,\theta,z,t)&=&-E_0\cdot\psi_{mn}(r,\theta).\cos\left(\frac{2\pi l}{L_z}z\right).\cos\left(\omega t\right)\non\\
\psi_{mn}(r,\theta)&=&C_{mn}\left(A_{mn}.J_n(\alpha_{mn}.r)+Y_n(\alpha_{mn}.r)\right)\times\non\\
&&\sin\left(\frac{2\pi n\theta}{\Theta}\right)\label{TM1}
\eea
with $l>0$ some integer and where $J_n(r),Y_n(r)$ are Bessel functions of the first and second kind, respectively. In the above equation, the constants $C_{mn}$ provide normalization, for instance they are such that $\max\Etm_z(r,\theta,z)=E_0$ in the cavity while the constants $A_{mn}$ and $\alpha_{mn}$ are given by the boundary condition $\Etm_z(r=R_{\rm in,out},\theta)=0$:
\bea
A_{mn}.J_n(\alpha_{mn}.R_{\rm in,out})+Y_n(\alpha_{mn}.R_{\rm in,out})&=&0\cdot
\eea
From those constants and the Helmholtz equation above, one has the following dispersion relation 
$$
\left(\frac{\omega_{mnl}}{c}\right)^2=\alpha_{mn}^2+\left(\frac{2\pi l}{L_z}\right)^2,
$$
where the wavenumber $k$ has been replaced by the value $2\pi l/L_z\cdot$
 The other components of the EM fields
in the TM polarization can be obtained from $\Etm_z$ as 
\bea
\vec{\Etm}_\perp&=& E_0 \frac{kc^2}{\omega^2-k^2c^2}\cdot\vec{\nabla}_\perp \psi_{mn}\cdot\sin\left(kz\right).\cos\left(\omega t\right)\label{TM2}\\
\vec{\Btm}_\perp&=&E_0 \frac{\omega}{\omega^2-k^2c^2}\cdot\vec{e}_z\wedge\vec{\nabla}_\perp \psi_{mn}\cdot\cos\left(kz\right).\sin\left(\omega t\right)\non
\eea
with $k=2\pi l/L_z$ and where a subscript $\perp$ still indicates transverse components (see also \cite{jackson}).
It can be verified that Eqs.(\ref{TM1},\ref{TM2}) are solutions of Maxwell equations with the appropriate boundary conditions on the surface of the hollow cavity. Eqs.(\ref{TM1},\ref{TM2}) can be combined with the assumption of uniform static external magnetic field $B_z=\rm cst$ to assemble the source terms
Eqs.(\ref{TmunuTM}). Under the assumptions of external longitudinal magnetic field $\vec{B}_\perp=0$, the GW polarizations that are excited by the TM cavity have the form, in the Lorenz gauge,
\be
h_{\mu\nu}^{\rm TM}=\left(
\begin{matrix}
0 & h_{01} & h_{02} & 0\\
h_{01} & 0 & 0 & h_{13}\\
h_{02} & 0 & 0 & h_{23}\\
0 & h_{13} & h_{23} & 0
\end{matrix}
\right)\cdot
\label{h_TM}
\ee

\begin{figure}[ht!]
\includegraphics[trim={0cm 0cm 0cm 0cm},clip,scale=0.5]{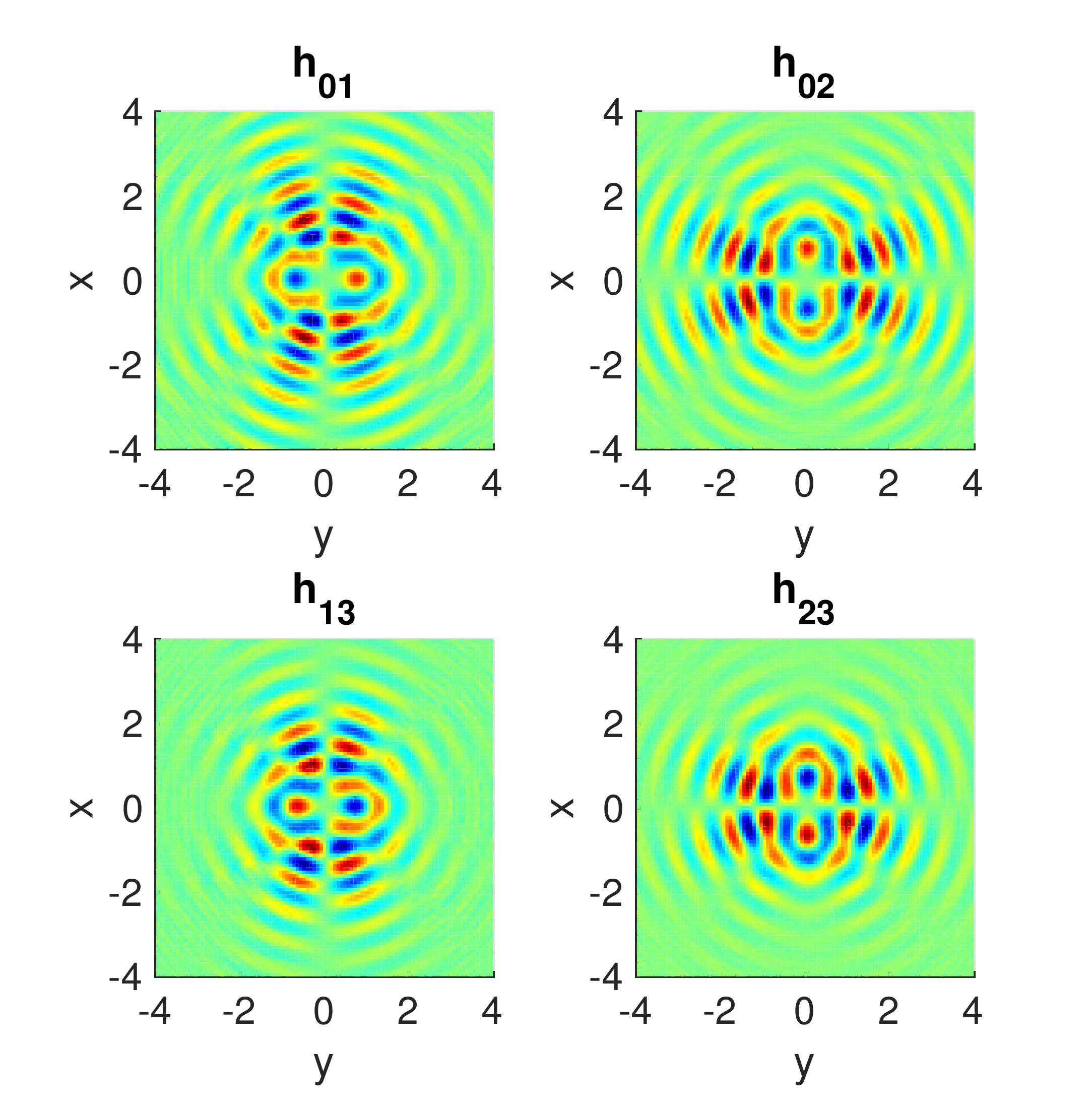}
\caption{Transverse sections $z=1$ of metric field perturbations $h_{01},h_{02},h_{13},h_{23}$
around a generator using TM wave resonance with external longitudinal magnetic field ($B_x=B_y=0$, $R_{\rm in}=0.2$, $R_{\rm out}=1$, $\Theta=\pi /2$ and $n=l=1\cdot$)}
\label{sections_hs_opencyl_trans}
\end{figure}

\begin{figure}[ht!]
\includegraphics[trim={0cm 0cm 0cm 0cm},clip,scale=0.5]{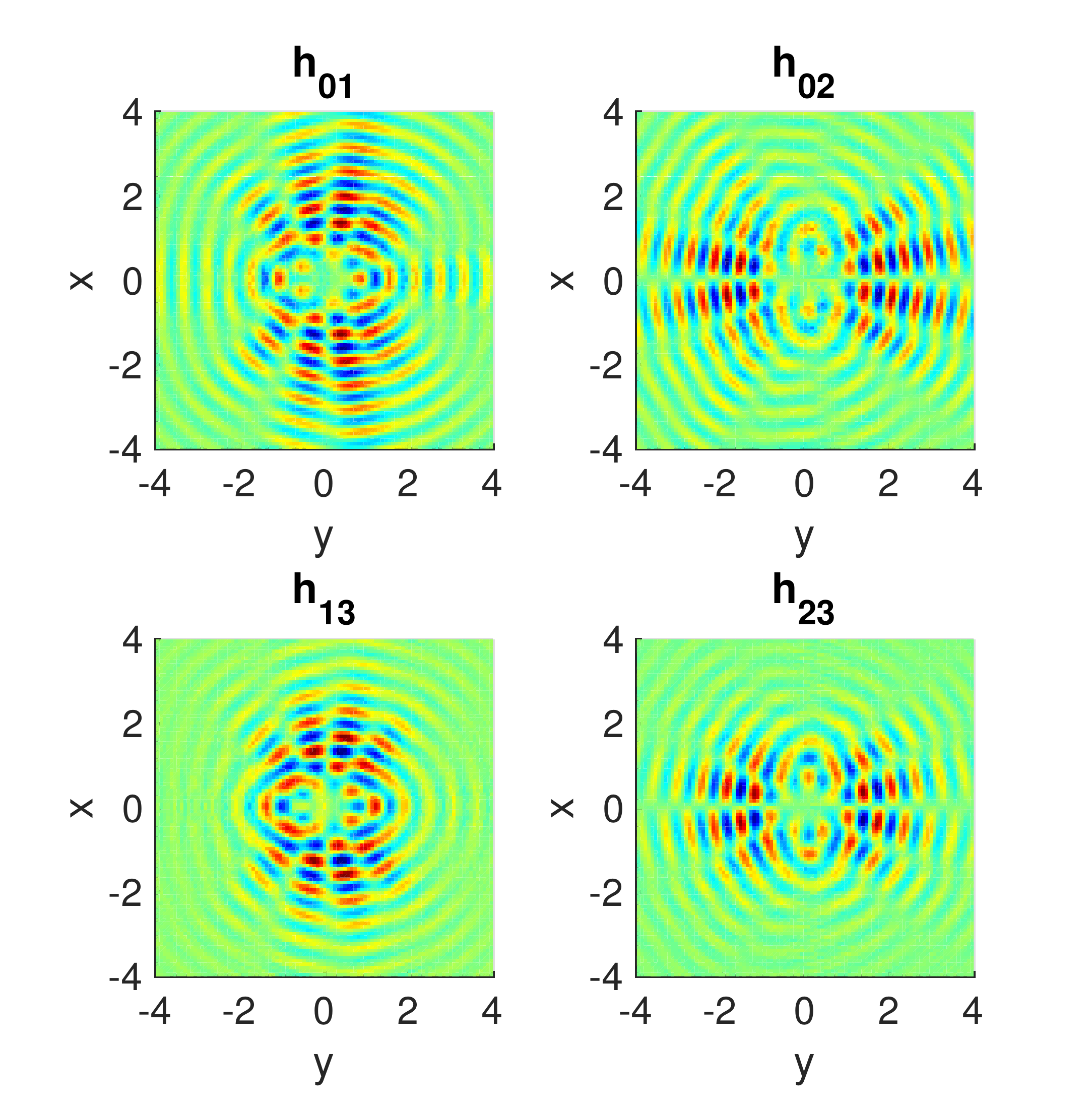}
\caption{Transverse sections $z=1$ of metric field perturbations $h_{01},h_{02},h_{13},h_{23}$
around a generator using TM wave resonance with external longitudinal magnetic field (same as above but with $\Theta=\pi /3$)}
\label{sections_hs_opencyl_transbis}
\end{figure}

\begin{figure}[ht!]
\includegraphics[trim={0cm 0cm 0cm 0cm},clip,scale=0.5]{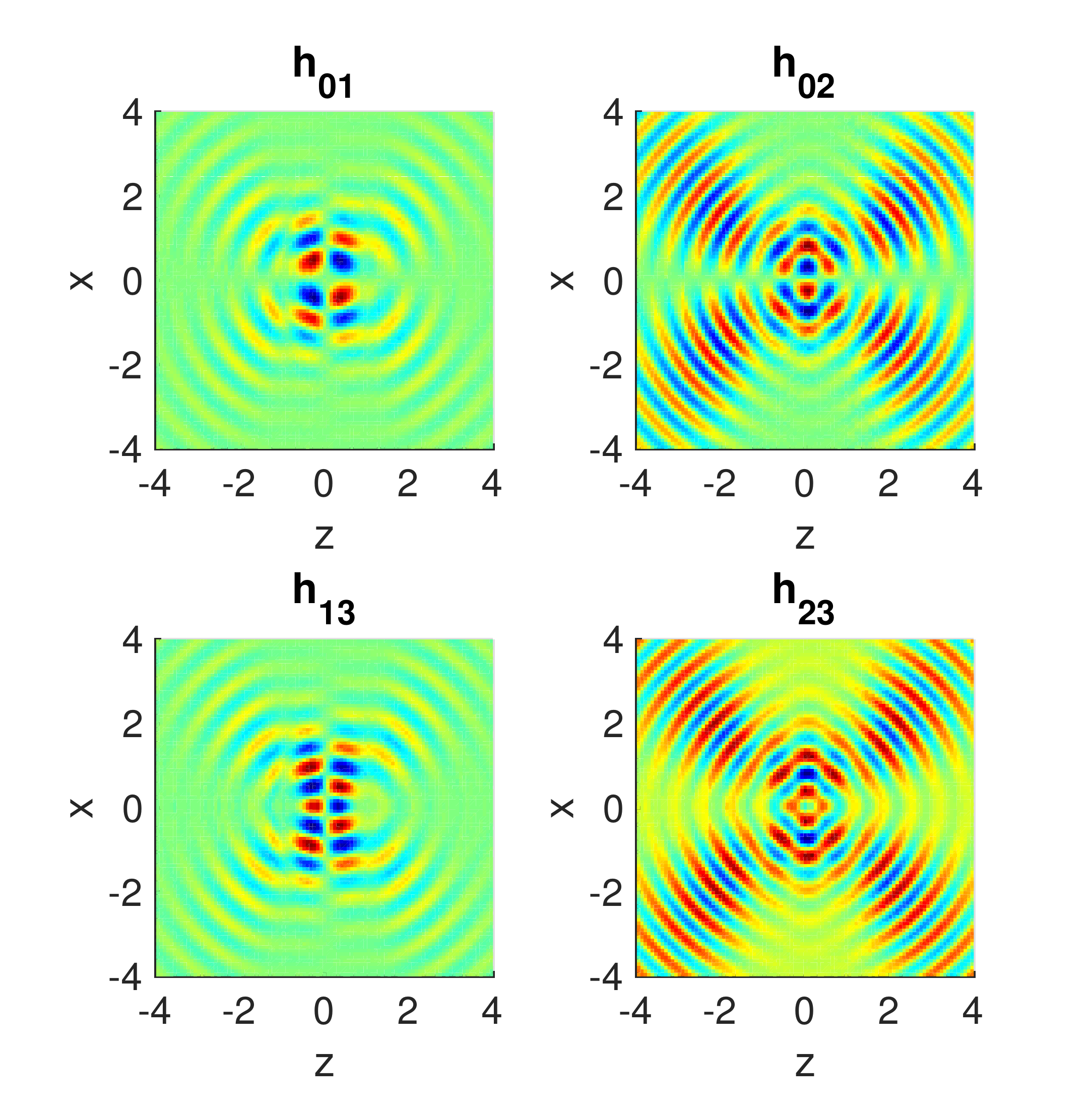}
\caption{Longitudinal sections $y=1$ of metric field perturbations $h_{01},h_{02},h_{13},h_{23}$
around a generator using TM wave resonance with external longitudinal magnetic field ($B_x=B_y=0$, $R_{\rm in}=0.2$, $R_{\rm out}=1$, $\Theta=\pi /2$ and $n=l=1\cdot$)}
\label{sections_hs_opencyl_longit}
\end{figure}

\begin{figure}[ht!]
\includegraphics[trim={0cm 0cm 0cm 0cm},clip,scale=0.5]{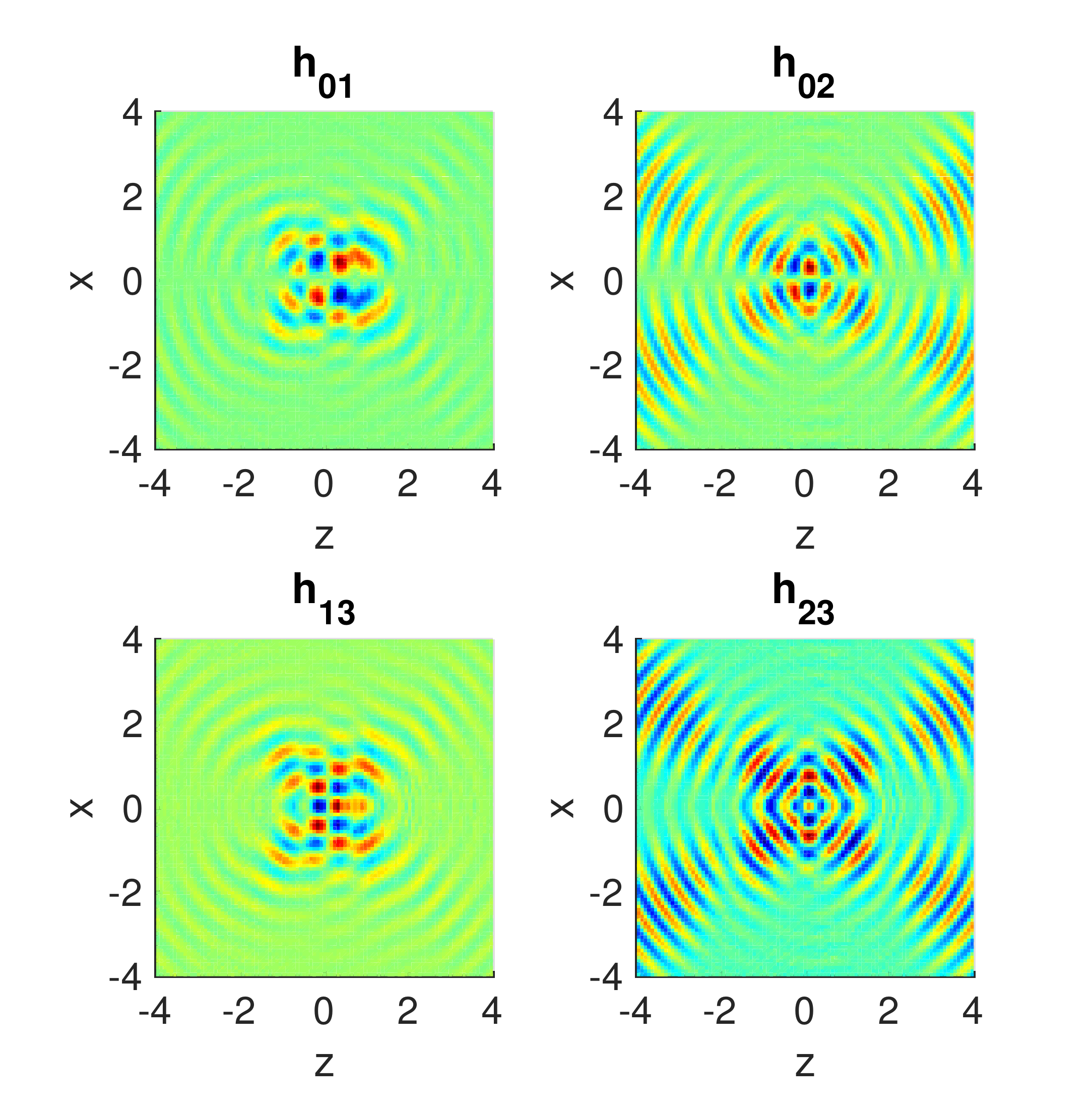}
\caption{Longitudinal sections $y=1$ of metric field perturbations $h_{01},h_{02},h_{13},h_{23}$
around a generator using TM wave resonance with external longitudinal magnetic field (same as above but with $\Theta=\pi /3$)}
\label{sections_hs_opencyl_longitbis}
\end{figure}

Figs. \ref{sections_hs_opencyl_trans}-\ref{sections_hs_opencyl_longitbis} represent transverse and longitudinal sections of metric perturbations $h_{01},h_{02},h_{13},h_{23}$ for device similar to the one presented in Fig.\ref{fig_gen_opencyl} for two different values of $\Theta$ (and for $R_{\rm in}=0.2$, $R_{\rm out}=1$, $n=l=1$ and $\Theta=\pi /2$ for Figs. \ref{sections_hs_opencyl_trans},\ref{sections_hs_opencyl_longit} and $\Theta=\pi/3$ for Fig. \ref{sections_hs_opencyl_transbis},\ref{sections_hs_opencyl_longitbis}). Besides exciting different GW polarizations, the generator presented here produces different radiation patterns and more focused directions of emission
compared to the generators presented before. This establishes that directed emission of GWs can be obtained with asymmetrical resonant cavities like the one in Fig.\ref{fig_gen_opencyl} with $\Theta<2\pi$. 
The amplitude of the emitted GW are of order of the Gertsenshtein-Zeldovich number Eq.(\ref{GZ}) up to some geometrical factor depending on the transverse dimensions of the GW generator.

It will be shown in section \ref{indirect} below that the outgoing energy radiated by GWs is also directional, and we will present there the gravitational radiation power patterns. 

Other GW polarizations can be excited by introducing a transverse component of the external magnetic field ($B_{x,y}\ne 0$). This allows emitting specific GW polarizations, that could also be detected by the technique proposed in the next section, which is sensitive to the incoming GW polarizations.

\clearpage
\section{Directional Reception of Gravitational Waves with Magnetic Energy Storage}
\label{detectors}
While several detection techniques based on EM properties have been proposed in the literature (see our introduction), we introduce now a novel method based on magnetic energy storage.

The principle of the method is as follows. Some GW passing through a static magnetic field $\vec{\Bdet}^{(0)}$ will locally modify the geometry, which results in a locally varying magnetic flux and the emission of EM waves $A^{(1)}_\mu$ (with $A_\mu$ the four-potential). This is sometimes dubbed the inverse Gertsenshtein effect\footnote{Although unfortunately discarded as being "\textit{hardly of interest}" by Gerstenshtein himself in his original paper.}: the conversion of GWs into EM ones we will reformulate below. The energy stored in the region covered by the magnetic field 
$\vec{\Bdet}^{(0)}$ is therefore modified by the passing GW. Indeed, the total magnetic field is the superposition of the static one $\vec{\Bdet}^{(0)}$ and the induced one $\vec{\Bdet}^{(1)}$ and the total EM energy density is quadratic in the norm of the total magnetic field $\vec{\Bdet}^{(0)}+\vec{\Bdet}^{(1)}\cdot$ 
When a GW induces an additional magnetic field in the magnet, the total energy stored by the magnetic field therefore varies by roughly an amount 
$$
\Delta E\sim \left(||\vec{\Bdet}^{(1)}||\cdot||\vec{\Bdet}^{(0)}||\right)\cdot \frac{V_{\rm det}}{\mu_0},
$$
with $V_{\rm det}$ the volume of the detector. Therefore, some large region of strong magnetic field can be used as a GW detector through monitoring precisely the amount of energy stored in the magnet. This could be done for instance using superconducting magnetic energy storage technology \cite{smes}, with special attention paid on the high-precision measurement of the discharge current when the unit is emptied. In addition, the passing GW interacts with the static magnetic field to produce new photons of the same frequency of the GW whose detection can be attempted. We will present below in \ref{planeGW} an estimation of the number of the photons generated through this process.

We can now examine the theory of the conversion of gravitational waves into EM ones.
In order to do this, we derive the inverse Gertsenshtein effect starting with the second group of covariant Maxwell equations\footnote{The first group, $\nabla_{\left[\alpha\right.} F_{\left.\beta\gamma\right]}=0$, being unchanged in the Levi-Civita connection.}
$\nabla_{\mu} F^{\mu\nu}=0$ perturbed at first order in the metric:
\bea
\nabla_{\mu} F^{\mu\nu}&\approx& \left(\eta^{\mu\alpha}\eta^{\nu\beta}-\eta^{\mu\alpha}h^{\nu\beta}-h^{\mu\alpha}\eta^{\nu\beta}\right)\partial_{\mu} F_{\alpha\beta}\non \\
&& -\eta^{\nu\beta}\partial_{\mu}h^{\mu\alpha} F_{\alpha\beta}-\eta^{\mu\alpha}\partial_{\mu}h^{\beta\nu} F_{\alpha\beta}\non\\
&&+\frac{1}{2}\partial_\mu h_\sigma^\sigma F^{\mu\nu}\cdot
\eea
If we now set $F_{\mu\nu}=F^{(0)}_{\mu\nu}+F^{(1)}_{\mu\nu}$ and if we focus solely on the first order correction $F_{\mu\nu}^{(1)}$, the previous equation simply reduces to
\be
\partial_{\mu} F_{(1)}^{\mu\nu}=-\eta^{\mu\alpha}\partial_{\mu}h^{\beta\nu} F^{(0)}_{\alpha\beta}\equiv \mathcal{J}^\nu\label{inverse_GZ}
\ee
neglecting higher order terms, given that $h_{\mu\nu}$ is traceless and satisfies the Lorenz gauge condition, and considering that
zeroth order Fadaray tensor components verifies Maxwell equation on Minkowski background $\partial_{\mu} F^{(0)}_{\alpha\beta}=0$. Equation (\ref{inverse_GZ}) shows us that the GWs $h_{\mu\nu}$ can combine with background EM fields $F_{(0)}^{\mu\nu}$ to constitute an effective 4-current density $j^\nu_{\rm eff}$ that generates weak EM fields $F_{(1)}^{\mu\nu}$.
We will solve Eq.(\ref{inverse_GZ}) under its wave form, by letting $F^{(1)}_{\mu\nu}=\pr_\mu A^{(1)}_{\nu}-\pr_\nu A^{(1)}_{\mu}$ and assuming Lorenz
gauge $\pr_\mu A_{(1)}^{\mu}=0$ such that
\be
\Box^2 A^{(1)}_{\mu}=\mathcal{J}_\mu\cdot\label{inverse_GZ2}
\ee
We will consider that the incoming GWs are passing by some region filled with a static uniform
magnetic field $\vec{\Bdet}^{(0)}$, and we will use the induced variation of the magnetic energy as a way to
detect GWs. 

From Eq.(\ref{inverse_GZ2}), we obtain that the induced EM fields
are given by
\be
A^{(1)}_{\mu}=\Bdet^{(0)}\cdot L_z\cdot \GZ\cdot a^{(1)}_{\mu}
\label{A1}
\ee
where $a^{(1)}_{\mu}$ is dimensionless and is solution of the
wave equation
\be
\Box_{(\vec{R},T)}^2 a^{(1)}_{\mu}=\textsl{J}_\mu
\label{inverse_GZ_adim}
\ee
with $\vec{R}^T=(x,y,z)/L_z=\vec{r}/L_z$, $T=ct/L_z$ and 
$$
\textsl{J}_\mu=\frac{ L_z}{||\vec{\Bdet}^{(0)}||\GZ}\cdot \mathcal{J}_\mu\cdot
$$
The source term in the above equation $\textsl{J}_\mu$
is of order unity from Eqs.(\ref{jeff_TEM},\ref{jeff_TM}), and so does $a^{(1)}_{\mu}$. The computation of the four-potential $a^{(1)}_{\mu}$ allows obtaining the additional electromagnetic field $F^{(1)}_{\mu\nu}$ that is induced in the detector.

We can now give the effective current densities that emerge when different GWs are passing by a region
of constant magnetic field. For incoming GWs in the traceless-transverse gauge (TT-gauge) or GWs generated by the TEM wave resonance mechanism, we can use the same coordinate system as in section II and consider the GW polarizations as in Eq.(\ref{h_TEM}), with vanishing longitudinal modes $h_{\fwd,\bwd}$ for the case of the TT gauge. The effective current densities in Eq.(\ref{inverse_GZ}) are then given by
\bea
\mathcal{J}_\fwd&=&\Bdet_y \partial_x h_\fwd-\Bdet_x \partial_y h_\fwd\label{jeff_TEM}\\
\mathcal{J}_\bwd&=&-\Bdet_y \partial_x h_\bwd+\Bdet_x \partial_y h_\bwd\non\\
\mathcal{J}^x&=&-\Bdet_y\partial_z h_{xx} + \Bdet_x \partial_z h_{xy}\non\\
\mathcal{J}^y&=&-\Bdet_y\partial_z h_{xy} - \Bdet_x \partial_z h_{xx}
\non
\eea
with $\mathcal{J}^\nu=\mu_0 j^\nu_{\rm eff}$, $\mathcal{J}^0=\mathcal{J}_\fwd+\mathcal{J}_\bwd$ and $\mathcal{J}^z=\mathcal{J}_\fwd-\mathcal{J}_\bwd\cdot$ In the above equation, we have used the fact that the Lorenz gauge condition  implies that $\partial_y h_{xx} -\partial_x h_{xy}=\partial_y h_{xy} +\partial_x h_{xx}=0\cdot$ From Eqs.(\ref{jeff_TEM}), one can therefore decompose the induced vector potential as following $A^{(1)}_t= A^{(1)}_\fwd+A^{(1)}_\bwd$, $A^{(1)}_z= A^{(1)}_\fwd-A^{(1)}_\bwd$ and choose an appropriate electromagnetic gauge in which $A^{(1)}_{t,z}$ both vanish. This shows that only transverse components $A^{(1)}_{x,y}$ are remotely induced
by the TEM wave resonance mechanism. This holds true also for  the case of incoming GWs in the TT gauge for which $h_{\fwd,\bwd}=0\cdot$

In the case where the incoming GWs are under the polarizations given by  Eq.(\ref{h_TM}) as generated by the TM wave resonance mechanism, the effective current densities in Eq.(\ref{inverse_GZ}) are given by
\bea
\mathcal{J}^{t}&=&\Bdet_z \left(\partial_x h_{02}-\partial_y h_{01}\right)+\Bdet_y \partial_z h_{01}-\Bdet_x \partial_z h_{02}\non\\
\mathcal{J}^x&=&\Bdet_y \partial_x h_{13}-\Bdet_x \partial_y h_{13}\non\\
\mathcal{J}^y&=&\Bdet_y \partial_x h_{23}-\Bdet_x \partial_y h_{23}\label{jeff_TM}\\
\mathcal{J}^z&=&\Bdet_z \left(\partial_y h_{13}-\partial_x h_{23}\right)-\Bdet_y \partial_z h_{13}+\Bdet_x \partial_z h_{23}
\non
\eea

With these tools in hand, one can consider some possible laboratory applications of electromagnetically generated or detected GWs.

\section{Applications: Experimental Concepts for GW Physics}
\label{applications}
In this section, we consider three new different GW physics experimental concepts and show that their detection threshold can be reached with present electromagnetic technology. Such experiments, in addition of being a premiere of gravity control in the lab, would allow testing general relativity in an unprecedented way, as we develop further in our conclusion below.
However, these concepts also face some expected technical challenges that we identify here in a non exhaustive way. These technical challenges should be investigated further before these concepts could be turned into real experiments. The results
presented in this section should therefore be considered as \textit{theoretical exploratory work} presented to motivate and provoke further work from the scientific community. 

\subsection{Indirect Detection of GW Emission}\label{indirect}
In section \ref{generators}, we have shown how EM resonant cavities put into external magnetic field produce GW with the same frequencies as the EM standing waves within the cavity. However, we focus there on the analysis of outgoing metric perturbations $h_{\mu\nu}$ which are not observables, since they are not gauge invariant. Observable quantities can although be built upon these metric perturbations and their derivatives.

Since fleeing GW carry energy and momentum away, this results in corresponding losses for the source. A first possible experimental application of our results is therefore the indirect detection of the emission of GWs from the measurement of the energy loss of the EM source. This power loss can be estimated from the energy-momentum tensor of the gravitational field $t_{\mu\nu}$ which, for traceless metric perturbations verifying Lorenz gauge condition such as those considered here, takes the form
\be
t_{\mu\nu}=\frac{c^4}{32\pi G} \langle \left(\partial_\mu h_{\alpha\beta}\right)\partial_\nu h^{\alpha\beta}\rangle
\label{tmunu_GW}
\ee
where $\langle\cdots\rangle$ denotes some average in space-time around the observing point located outside of the GWs source (see for instance \cite{hobson}). 

Let us first estimate the energy loss in the generator due to the emission of GWs. The loss corresponds to the energy carried away by the fleeing GWs (see also \cite{peres})  which we estimate by computing the radial energy flux of these waves over a sphere large enough to encompass the GW generator \cite{hobson}:
\be
\frac{dE}{dt}=\frac{c^4r^2}{32\pi G}\int_{4\pi}  \langle \left(\partial_t h_{\alpha\beta}\right)\partial_r h^{\alpha\beta}\rangle d\Omega\label{dEdt}
\ee
with $r$ denoting the radius of the encompassing sphere. If we consider $r\sim L_z$ and $\langle \left(\partial_t h_{\alpha\beta}\right)\partial_r h^{\alpha\beta}\rangle\sim \GZ^2c/L_z^2$ (with $L_z$ the characteristic length of the generator), we find the following order of magnitude for the power loss by emission of GWs:
\be
\left|\frac{dE}{dt}\right|\sim \frac{G E_0^2B_0^2 L_z^4}{c^5\mu_0^2}\cdot
\ee
If we consider a generator using petawatt laser pulses of intensity $E_0\sim 10^{15}$V/m and length $L_z\sim 10^{-6}$m into intense pulsed magnetic fields of $B_0\sim 100$T, one gets that the power loss is of order $10^{-31}$W per laser shot. In addition, the duration of the GW emission, typically
the traveling time of the pulse into the external magnetic field, is very short. Therefore, this seems very hard to detect. It seems better to adopt another strategy by using a continuous emission of GW, which can be carried out in a generator constituted of a standing wave immersed into a static intense magnetic field. A generator of size $L=10\rm m$ containing a standing wave of peak electric field $E_0=10^6$V/m put into an external magnetic field $B_0$ of $10$T will produce GWs of amplitude of order $\GZ=10^{-39}$ and will undergo an instantaneous power loss of $10^{-23}$W due to the fleeing GWs.
It is worth noticing that the similar experimental design of a microwave cavity in an external static magnetic field used in the ADMX experiment searching for axions has a sensitivity of $10^{-24}$W over a time integration of $10^3$s \cite{ADMX}. 
However, the situation for the indirect detection of GWs is quite different since it requires to measure a tiny variation of power inside a resonant EM cavity filled with high amplitude standing waves while in axion searches, one searches for a comparable tiny power gain but on top of vacuum. Future works should investigate whether one can successfully apply these detection techniques into the high energy environment of a filled cavity. 

\begin{figure}[ht!]
\begin{tabular}{cc}
\includegraphics[trim={10cm 5cm 13cm 5cm},clip,scale=0.2] {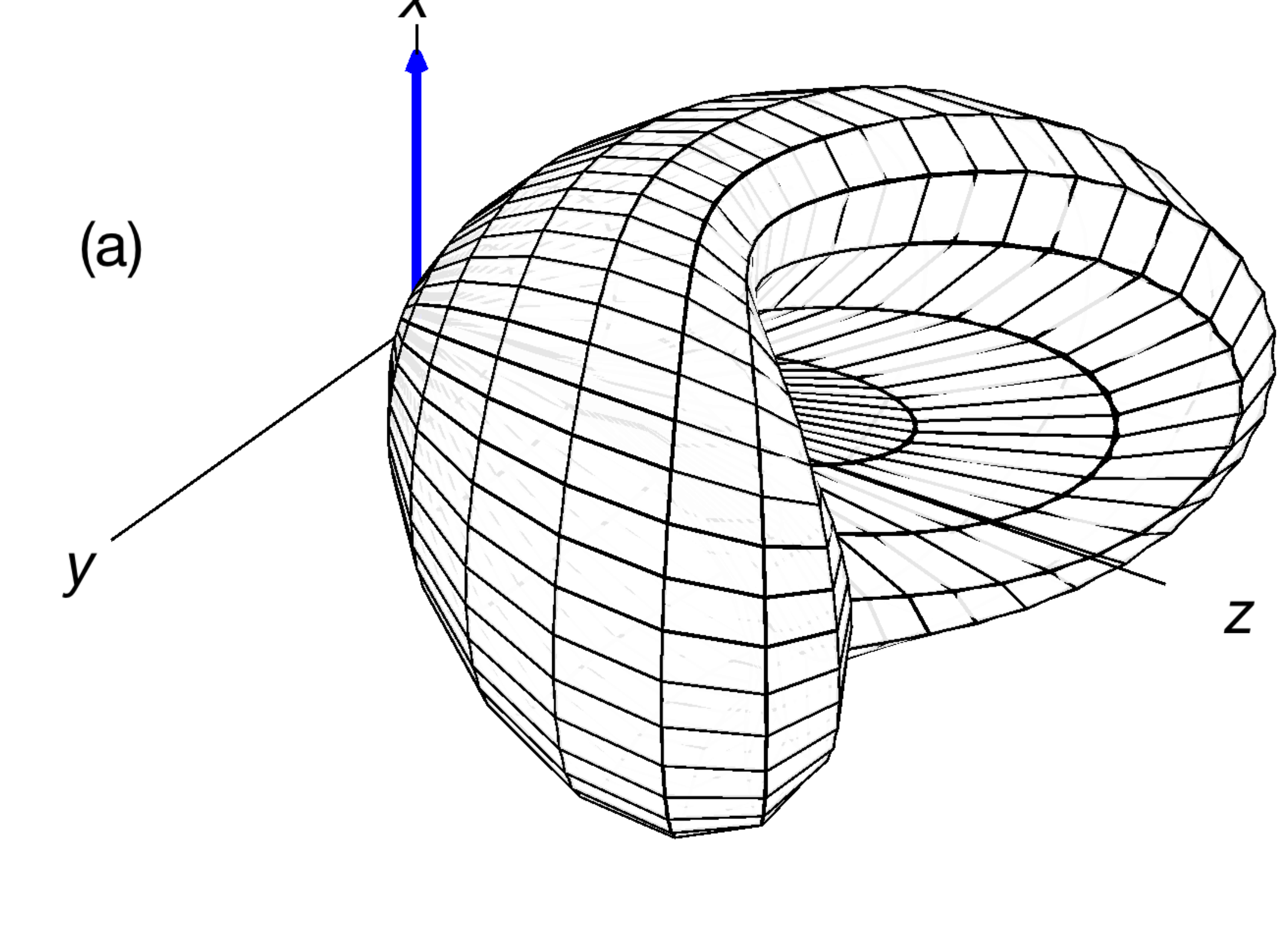}&
\includegraphics[trim={10cm 5cm 13cm 5cm},clip,scale=0.2] {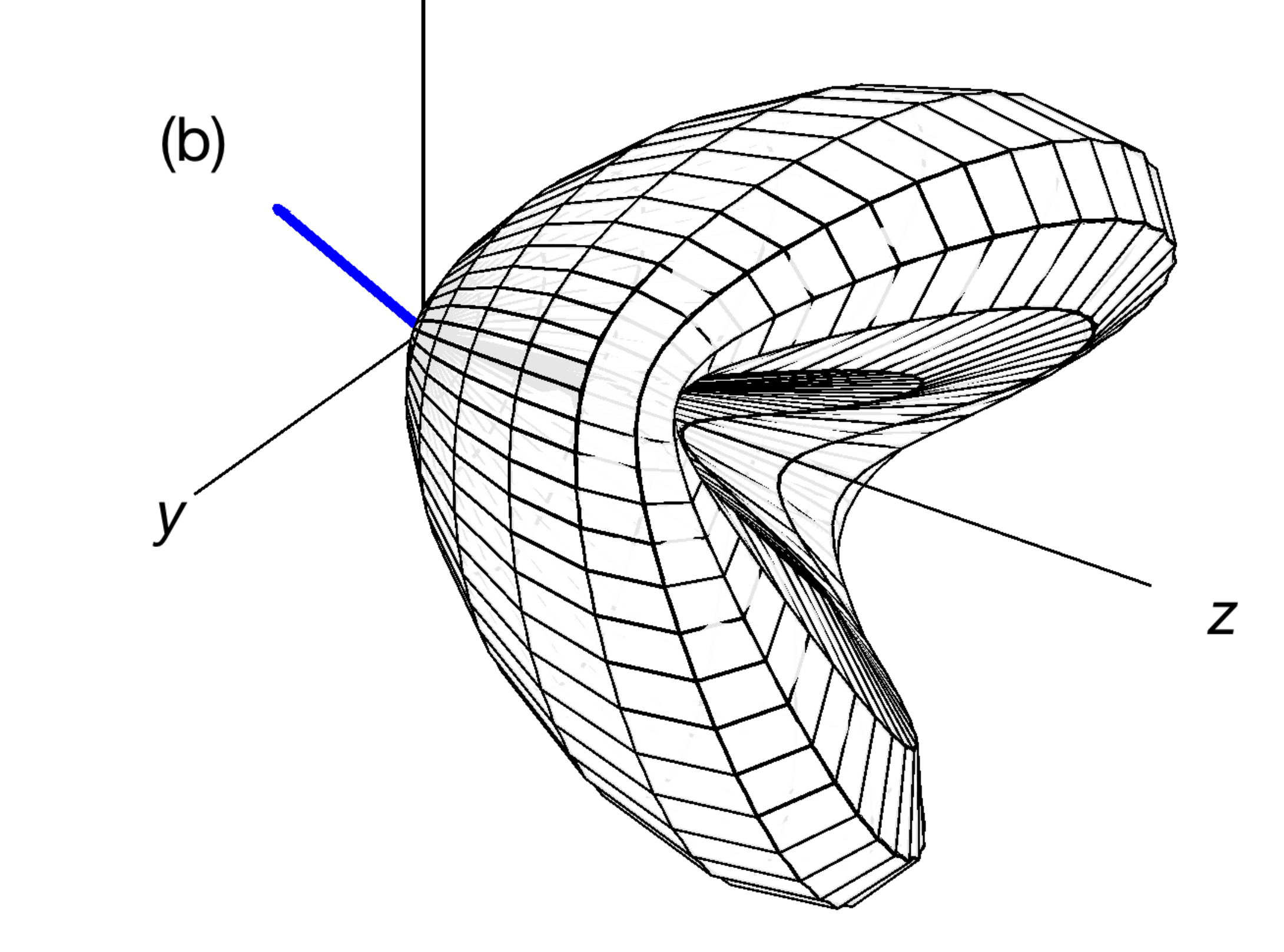}\\
\includegraphics[trim={10cm 5cm 13cm 5cm},clip,scale=0.2] {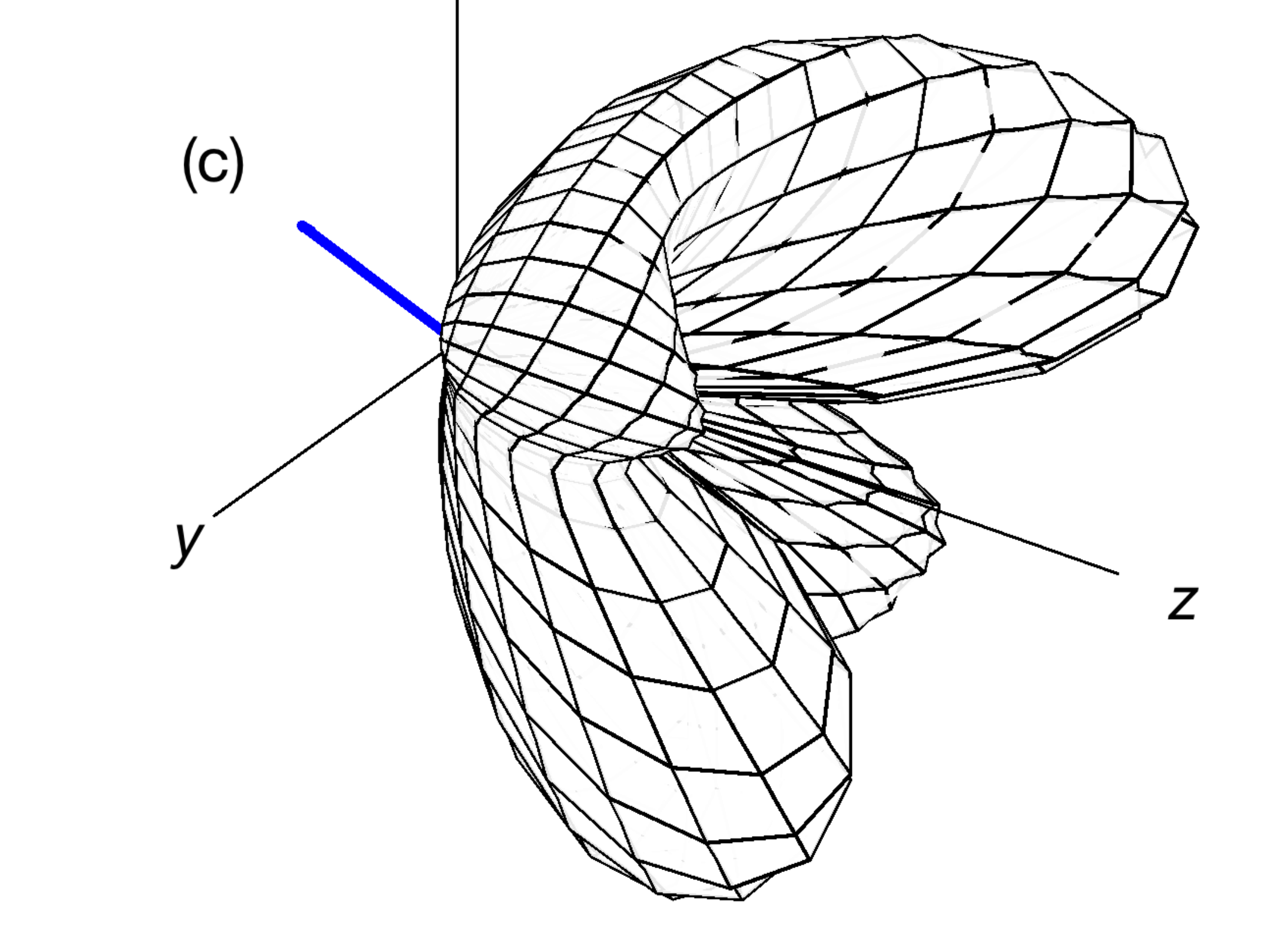}&
\includegraphics[trim={10cm 5cm 13cm 5cm},clip,scale=0.2] {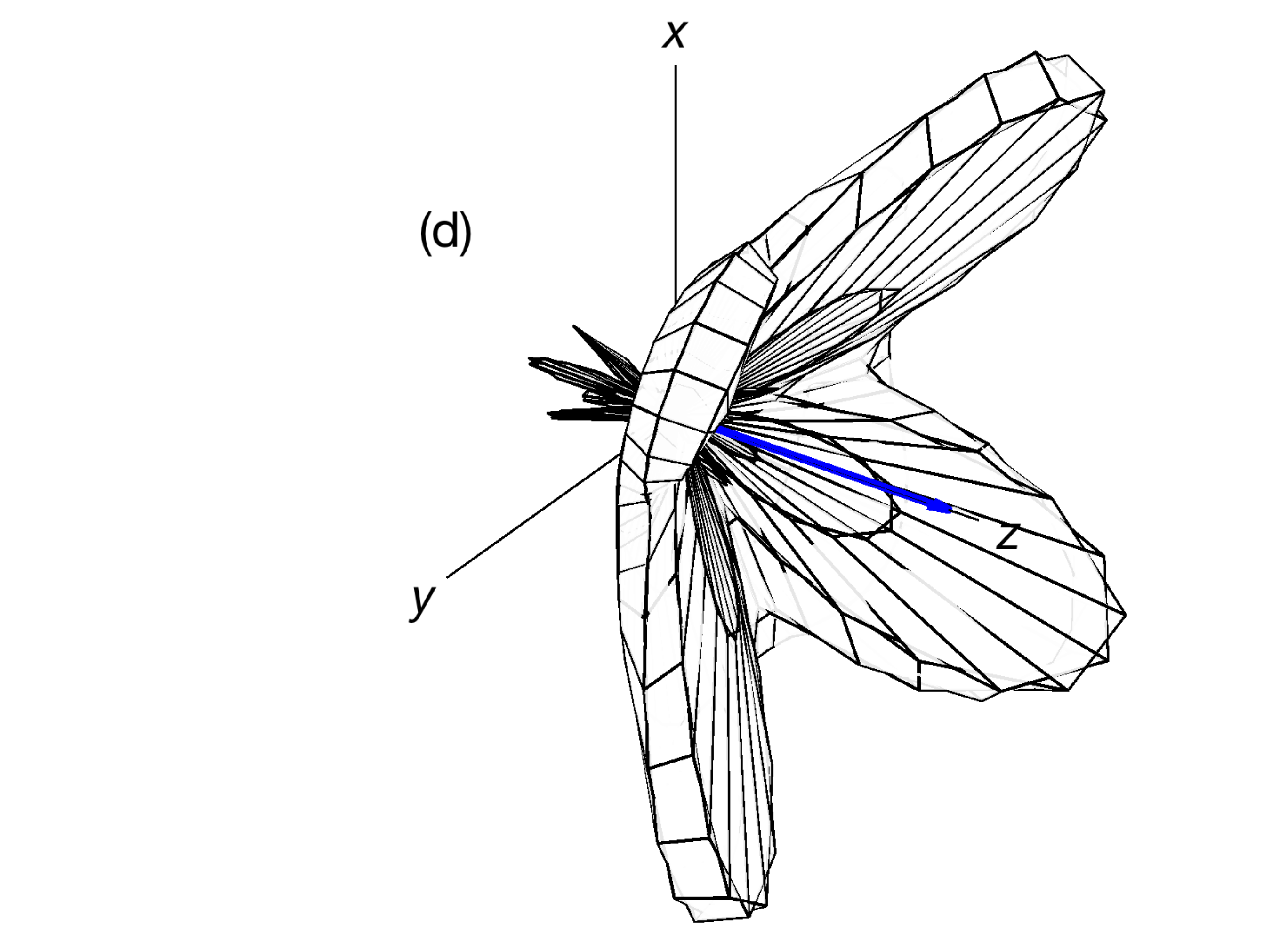}
\end{tabular}
\caption{Gravitational radiation power patterns: (a) TEM wave resonance generator with cylindrical waveguide ($R_{\rm in}=0.1$, $R_{\rm out}=1$, $\vec{B}_{\rm ext}=B_0\vec{e}_x$); (b) TEM wave resonance generator with rectangular waveguide ($L_x=L_y=L_z=1$, $L_x/l_x=L_y/l_y=3$, $\vec{B}_{\rm ext}=B_0\left(\frac{\sqrt{2}}{2} \vec{e}_x+\frac{\sqrt{2}}{2} \vec{e}_y\right)$); (c) TEM wave resonance generator with rectangular waveguide ($L_x=3$, all other parameters being the same as subpanel (b));
(d) TM wave resonance generator with open cylindrical cavity ($R_{\rm in}=0.5$, $R_{\rm out}=1$, $\Theta=\pi$, $\vec{B}_{\rm ext}=B_0\vec{e}_z$). Blue arrows indicate the external magnetic field $\vec{B}_{\rm ext}$}
\label{GWradpattern}
\end{figure}
\bigskip
Let us now turn on the directivity of the GW energy flux. Energy is obviously not radiated isotropically from our generators and we can use Eqs.(\ref{tmunu_GW}-\ref{dEdt}) to compute the gravitational radiation power pattern, in the same way this is done for a true electromagnetic antenna \cite{balanis}. To do this, we compute the gravitational radiation intensity Eq.(\ref{dEdt}) on a sphere of large radius to obtain the angular distribution of gravitational energy radiation. The energy fluxes 
$t_{0\nu}$ in Eq.(\ref{tmunu_GW}) are obtained by computing $\langle\cdots\rangle$ as the time average over one period, since the EM standing waves used in the generators are pure harmonic modes of pulsation $\omega$ that subsequently produce monochromatic GWs of same frequency. The gravitational radiation power patterns in the region of $z>0$ for different generators are given in Fig. \ref{GWradpattern}. 

The GW emission from our EM devices is clearly directional and quadrupolar as expected. For TEM wave resonance generators, the directions of maximal GWs emission are orthogonal to the orientation of the external magnetic field (in blue in Fig. \ref{GWradpattern}) and the emission lobes develop along the axis of the TEM waveguide. In order to change the direction of maximum GW emission, one has to rotate the transverse external static magnetic when using a generator with cylindrical TEM waveguide. For a generator with a TEM waveguide with rectangular cross-sections, one has simply to rotate the rectangular waveguide inside the external magnetic field to change the direction of maximal GW emission. This gives a clear practical advantage for the latter design. Increasing the transverse length of the TEM waveguide, as is done in the subpanel (c) of Fig. \ref{GWradpattern} deforms the lobes of emission, with an increase of the radiation intensity in the direction of the longer transverse dimension. The gravitational radiation power pattern of a generator with a TM half-cylinder hollow cavity ($\Theta=\pi$) in a longitudinal static magnetic field is illustrated
in the subpanel (d) of Fig. \ref{GWradpattern}: it shows four major 
 emission lobes and four minor lobes around the direction of the external longitudinal magnetic field. The number of lobes, their orientation and their thickness all depend on the aperture $\Theta$ of the open-cylindrical TM cavity.

\subsection{Direct Detection of Incoming GWs With Strong Magnetic Field}\label{planeGW}
Let us now consider an incoming plane GW in the TT-gauge propagating along the $z-$direction
\be
h_{\mu\nu}^{\rm TT}=\left(
\begin{matrix}
0 & 0 & 0 & 0\\
0 & h_{+} & h_{\times} & 0\\
0 & h_{\times} & -h_{+} & 0\\
0 & 0 & 0 & 0
\end{matrix}
\right)
\label{h_TT}
\ee
with
$$
h_{+,\times}=\mathcal{H}_{+,\times}\cos\left(kz-\omega t\right)
$$
where $\mathcal{H}_{+,\times}$ is the amplitude of the related GW polarization and where $k=\omega/c$
is the gravitational wavenumber ($\omega=2\pi c/\lambda_{\rm GW}$, $\lambda_{\rm GW}$ being the GW wavelength). This plane GW then passes through some region of static magnetic field $\vec{\Bdet}^{(0)}$ where it induces an extra EM potential $A^{(1)}_\mu$ as a consequence of the interaction between the GW and the magnetic field in the detector. 

Let us denote by $\vec{\Bdet}^{(0)}$ the magnetic field of the detector. From Eq.(\ref{inverse_GZ}), we get the following effective current densities that source  $A^{(1)}_\mu$ into the 
wave equation Eq.(\ref{inverse_GZ2}):
\bea
\mathcal{J}^x&=&-\Bdet^{(0)}_y\partial_z h_{+} + \Bdet^{(0)}_x \partial_z h_{\times}+ \Bdet^{(0)}_z\left( \partial_y h_{+}-\partial_x h_{\times}\right)\non\\
\mathcal{J}^y&=&-\Bdet^{(0)}_y\partial_z h_{+} - \Bdet^{(0)}_x \partial_z h_{\times}+ \Bdet^{(0)}_z\left( \partial_y h_{\times}+\partial_x h_{+}\right)\cdot\label{jeff_TT}
\non
\eea
Under the assumptions of a plane wave above and of a static magnetic field $||\vec{\Bdet}^{(0)}||=\Bdet_0$, Eq.(\ref{inverse_GZ_adim}) can easily be solved analytically. Furthermore, we have that $|\partial_{x,y} h_{+,\times}|\ll |\partial_{z} h_{+,\times}|$ for a plane wave so that only transverse components of the magnetic field $\Bdet^{(0)}_{x,y}$ produce the fields $A^{(1)}_\mu$ that have to be detected. Hence, this detection method is directional since only plane GWs coming from a direction perpendicular to
the static magnetic field $\vec{\Bdet}^{(0)}$ will produce a signal. Without loss of generality, we can consider that the x-axis is aligned along $\vec{\Bdet}^{(0)}$.
\\
\\
We propose to detect the incoming plane GWs through the variation of the magnetic energy stored in the detector. Neglecting the contribution of the induced electric field $E^{(1)}$, the magnetic energy stored in the volume
of detection $V_{\rm det}$ is given by
\be
E=\frac{1}{2\mu_0}\int_{\rm det} ||\vec{\Bdet}_{\rm tot}||^2 dV 
\ee
where the total magnetic field in the detector $\vec{\Bdet}_{\rm tot}$ is the superposition of $\vec{\Bdet}^{(0)}$ and the magnetic field $\vec{\Bdet}^{(1)}$ induced by the passing GWs: $\vec{\Bdet}_{\rm tot}=\vec{\Bdet}^{(0)}+\vec{\Bdet}^{(1)}$. The absolute variation of magnetic energy in the detector
when the GWs is passing is given by
\bea
\Delta E&=&E\left(\vec{\Bdet}^{(0)}+\vec{\Bdet}^{(1)}\right)-E\left(\vec{\Bdet}^{(0)}\right)\non\\
&\approx& \frac{1}{\mu_0}\int_{\rm det}\vec{\Bdet}^{(0)}\bullet\vec{\Bdet}^{(1)}dV,
\label{dE}
\eea
at first order in $||\vec{\Bdet}^{(1)}||$.

Since $\vec{\Bdet}^{(1)}=\vec{\nabla}\wedge\vec{A}^{(1)}$ and given our assumption that $\vec{\Bdet}^{(0)}=\Bdet_0 \vec{e}_x$,
we only need to consider $\Bdet^{(1)}_x=\partial_z A^{(1)}_y$ and integrate it over the detection volume to get the absolute variation of energy $\Delta E$. We can consider for simplicity that the detector is a simple cylindrical solenoid of radius $R_{\rm det}$ and length
$L_{\rm det}$. Then, from the analytical solution of Eqs.(\ref{inverse_GZ_adim},\ref{jeff_TT}) with
the above assumptions, one can obtain the following expression for the variation of the stored energy in the detector
\be
\Delta E(t)\approx \frac{\pi c\Bdet_0^2 R_{\rm det}^2 \mathcal{H}_+}{\mu_0}\cdot t\cdot\sin\left(\frac{\pi L_{\rm det}}{\lambda_{\rm GW}}\right)\cdot \sin\left(\frac{2\pi c.t}{\lambda_{\rm GW}}\right)\cdot
\label{dE_plane}
\ee
The amplitude of the energy variation is therefore increasing with the total duration $t$ of the GW train passing through
the magnetic field region. For a GW signal with amplitude $\mathcal{H}_+\approx 10^{-21}$ and a detector of radius $R_{\rm det}=1$m producing a magnetic field of $\Bdet_0=10$T, we get an amplitude of energy variation of $\Delta E(t)\approx 10^{-4}\times t \times \sin\left(\pi L_{\rm det}/\lambda_{\rm GW}\right)$ Joules.  
\\
\\
In summary, one can consider using magnetic energy storage systems as GW detectors provided 
the amount of energy stored in the system can be very precisely monitored. This could be done through
the precise measurement of the outgoing electric current when the unit is emptied, the total magnetic field inside the unit or even the fluctuations in some EM or photon detector placed inside the magnetic field. Indeed, GWs passing through a constant magnetic field produce 
EM fields and hence photons of the same frequency. However, this requires to rely on high-precision measurement techniques for huge electric currents and magnetic fields, or high-sensitivity photon detection
in strong magnetic field environment, which can be expected to be technically quite challenging. 

\subsection{GW Emission-Reception Experiments}\label{HertzGW}

\begin{figure}[ht!]
\includegraphics[trim={4cm 4cm 4cm 3cm},clip,scale=0.3]{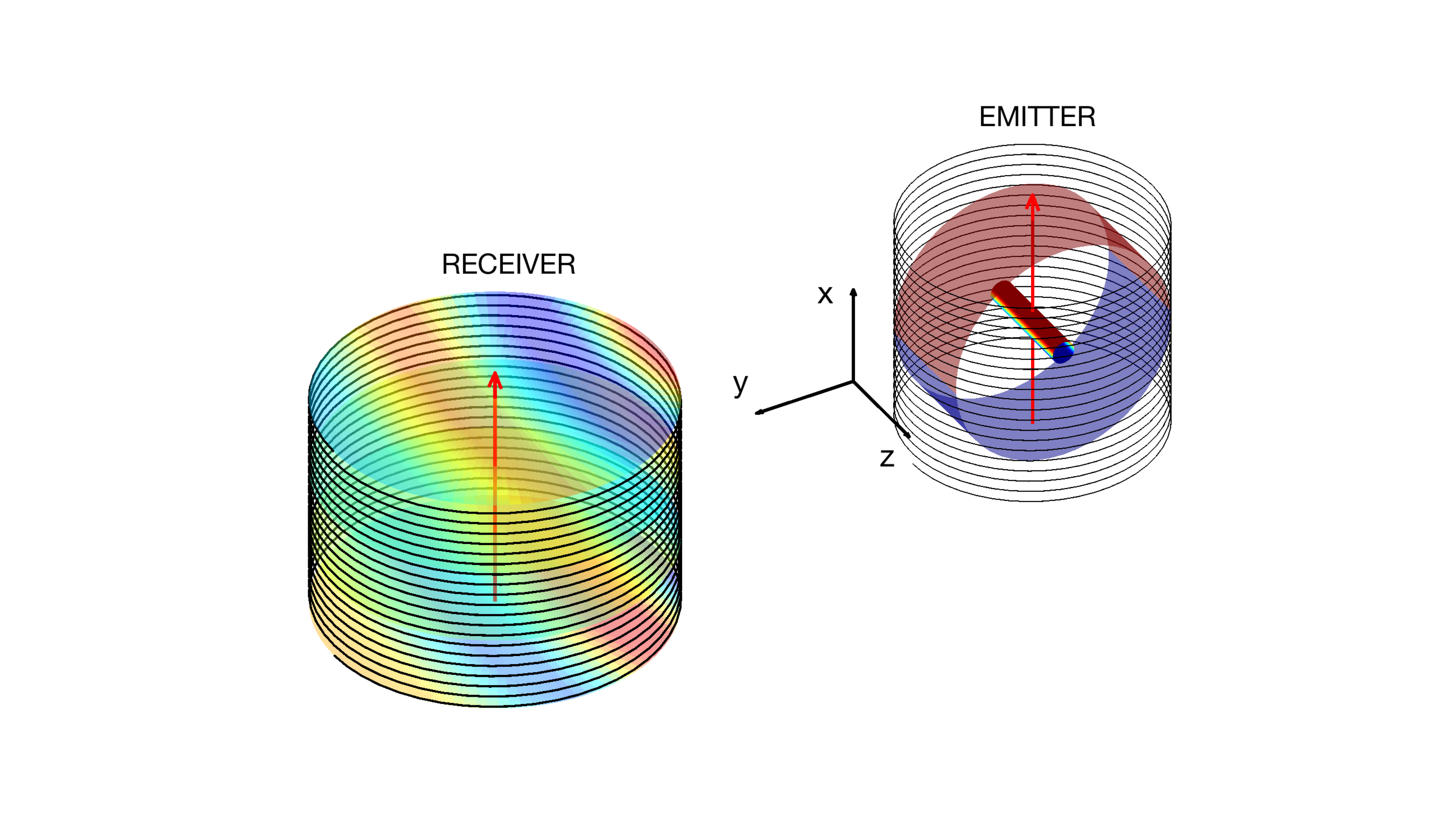}
\caption{Proposal for a GWs emission-reception experiment using TEM wave resonance. Red arrows illustrate the magnetic fields in the emitter and the receiver. Colored regions illustrate the GWs passing through the receiver}
\label{fig_exp_TEMres}
\end{figure}

\begin{figure}[ht!]
\includegraphics[trim={10cm 5cm 0cm 3cm},clip,scale=0.3]{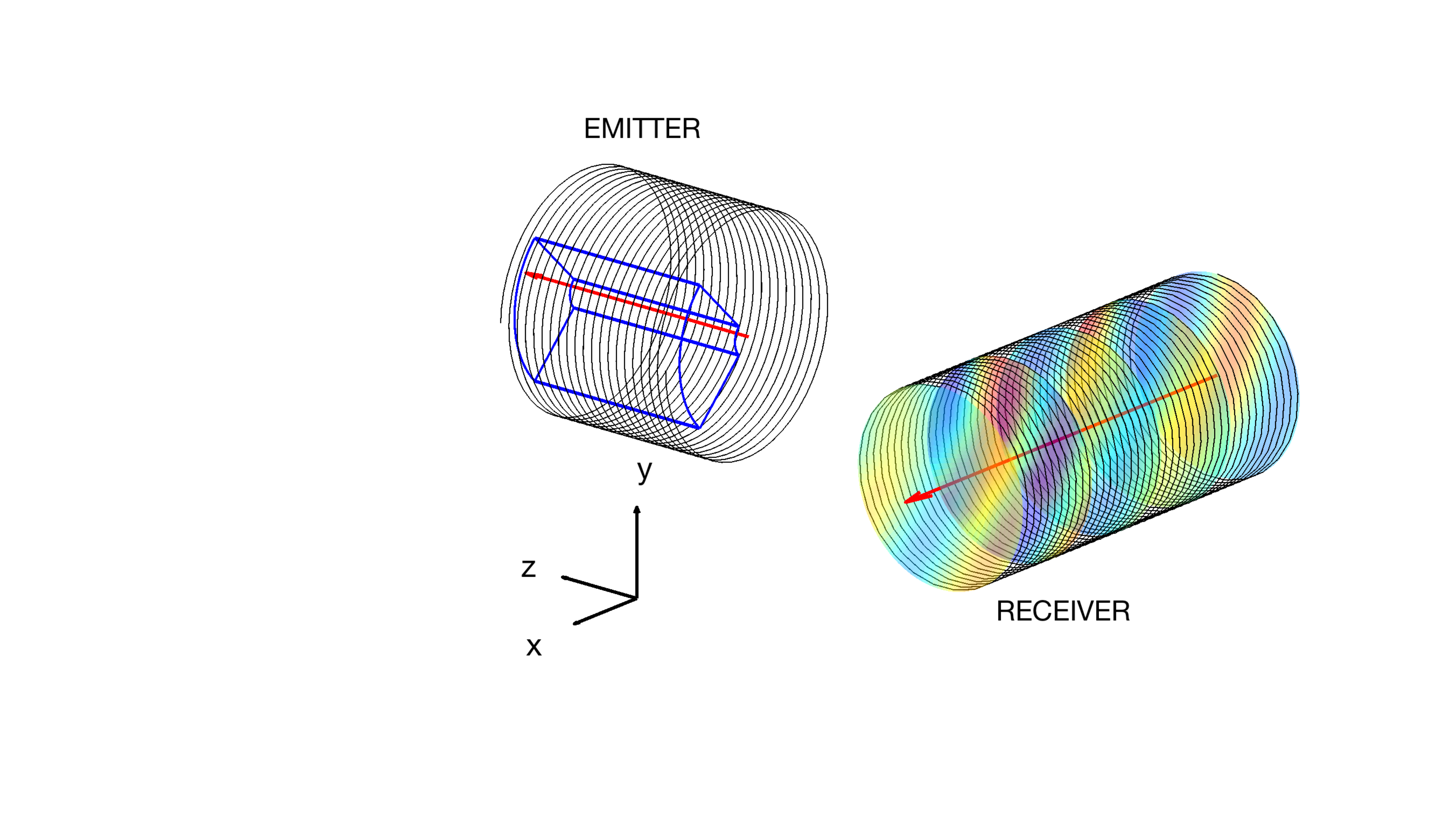}
\caption{Proposal for a GWs emission-reception experiment using TM wave resonance. Red arrows illustrate the magnetic fields in the emitter and the receiver. Colored regions illustrate the GWs passing through the receiver}
\label{fig_exp_TMres}
\end{figure}

We can now consider the even harder experiment involving GW emission and reception. The emitter will be given by one of the designs presented in section \ref{generators}, either using TEM or TM wave resonance. As for the receiver, we simply consider a remote region filled by a static magnetic field $\vec{\Bdet}$, like the interior of a solenoid, in which the GWs generated by the emitter will induce EM fields $A^{(1)}_\mu$. We assume the magnetic field $\vec{\Bdet}^{(0)}$ is constant over
the volume of the receiver for the sake of simplicity and is aligned so that a specific GW polarization can be detected, as we show below. There is no resonant cavity in the receiver considered here.
\\
Figs.\ref{fig_exp_TEMres} and \ref{fig_exp_TMres} illustrate two experimental concepts, with specific alignments of the magnetic fields in the emitter and the receiver.

As in the section \ref{planeGW}, we propose to detect the GWs emitted by the generator through the variation of the magnetic energy stored in the receiver whose order of magnitude is given by
\bea
\Delta E&\approx &\frac{||\vec{\Bdet}^{(0)}||}{\mu_0}\int_{\rm det}||\vec{\Bdet}^{(1)}||dV\cdot
\label{dE}
\eea

The absolute variation of stored magnetic energy is therefore proportional to the average of the induced magnetic field $||\vec{\Bdet}^{(1)}||$ over the volume of detection. Since $\vec{\Bdet}^{(1)}=\vec{\nabla}\wedge\vec{A}^{(1)}$ and 
considering Eqs.(\ref{jeff_TEM}-\ref{A1}), we have that 
$||\vec{\Bdet}^{(1)}||\sim ||\vec{\Bdet}^{(0)}||\GZ$ (up to the factor $a_{\mu}\sim (L_z/d)$ with $d$ the distance from the emitter
and $L_z$ the characteristic size of the generator). We finally find that the absolute variation in the stored magnetic energy is of rough order of magnitude:
\be
\Delta E\sim \frac{\left(\Bdet^{(0)}\right)^2V_{\rm det}\GZ}{\mu_0}
\label{dEbis}
\ee
up to the factor $L_z/d$ giving the distance between the emitter and the receiver. This estimation applies to a separation $d$ of one generator length $L_z$, giving some room for experimental equipment surrounding the emitter and the receiver such as the magnet return paths, etc..

Since the induced EM fields $A_\mu^{(1)}$ have the same frequency $\omega_{\rm GW}$ as the passing GWs (see Eqs.(\ref{inverse_GZ2}-\ref{jeff_TM})), the number of induced photons in the receiver is given by
\be
N_\gamma=\frac{\Delta E}{\hbar \omega}\sim \frac{\left(\Bdet^{(0)}\right)^2V_{\rm det}\GZ L_z}{2\pi c\mu_0\hbar}\label{Ngam}
\ee
since $\omega=2\pi c/\lambda_{\rm GW}\sim 2\pi c/L_z\cdot$ If we assume $V_{\rm det}\approx L_z^3$, the number of induced photons $N_\gamma$ in the receiver is of order $1$ for $L_z=10$m, $E_0=10^6$V/m, $B_0=\Bdet^{(0)}=14$T by using Eq.(\ref{GZ}).

The reasoning detailed just above supposes the average of $\vec{\Bdet}^{(1)}$
is constant over the volume of the receiver and its amplitude is of order of magnitude $\Bdet^{(0)}\GZ$. In practice, this average varies with time and the receiver is located some distance $d$ away from the emitter, so that the local amplitude of the passing GWs is actually smaller, of order $\GZ L_z/d$. 

To be more precise, let us now present the results of a full numerical computation that uses the GWs generated by the emitters presented in section II and the numerical resolution of the 3D wave equation Eq.(\ref{inverse_GZ_adim}) for the induced EM field $A^{(1)}_\mu$ in a distant receiver. The wave equation Eq.(\ref{inverse_GZ_adim}) will be solved for some volume filled with a uniform magnetic field $\vec{\Bdet}$ by using a spectral method based on Fourier Transform. More precisely, if we expand
$a^{(1)}_{\mu}$ over a basis of plane waves as
$$
a^{(1)}_{\mu}(\vec{R},T)\approx\sum_{\vec{K}} \hat{a}_{\mu,\vec{K}}(T)\exp\left(i\vec{K}\cdot\vec{R}\right)
$$
with $\hat{a}_{\mu,\vec{K}}(T)$ the Fourier transform of the field $a^{(1)}_{\mu}$ (which can be approximated for efficiency by a Fast Fourier Transform), then the wave equation Eq.(\ref{inverse_GZ_adim}) becomes a set of ordinary differential equations describing each Fourier coefficient as a forced harmonic oscillator:
$$
\frac{d^2 \hat{a}_{\mu,\vec{K}}(T)}{dT^2}+K^2\hat{a}_{\mu,\vec{K}}(T)=\hat{\textsl{J}}_{\mu,\vec{K}}(T)
$$ 
where $\hat{J}_{\mu,\vec{K}}(T)$ is the Fourier transform of the source term $\textsl{J}_\mu\cdot$ Those source terms are related to the gradients of the metric perturbations (see Eqs.(\ref{jeff_TEM},\ref{jeff_TM})) and can be obtained by differentiating Eq.(\ref{sol_h}). As a matter of boundary conditions, we will assume a simple case where the magnetic field $\vec{\Bdet}$ used for detection is switched on at time $T=0$, so that $\hat{a}_{\mu,\vec{K}}(T=0)=\left(d\hat{a}_{\mu,\vec{K}}/dT\right)_{T=0}=0$ and $\hat{\textsl{J}}_{\mu,\vec{K}}(T<0)=0\cdot$

We consider here two possible experimental realizations for these simulations. The first one (a)  (see Fig. \ref{fig_exp_TEMres}) involves a generator made of a cylindrical TEM waveguide in a transverse magnetic field, 
with parameters $R_{\rm in}=0.2$, $R_{\rm out}=1$ and $\lambda=L_z/2$. 
In the second one (b) (see Fig. \ref{fig_exp_TMres}), we consider the open cylinder TM cavity in a longitudinal magnetic field with parameters $R_{\rm in}=0.2$, $R_{\rm out}=1$, $\Theta=\pi/2$, $m=n=l=1$. In both cases (a-b), the receiver is constituted by some volume filled with a constant magnetic field $\vec{\Bdet}$ aligned along the $x$-axis. The receiver is put at a good distance from the emitter (see also Figs. \ref{fig_exp_TEMres} and \ref{fig_exp_TMres}) and not in contact with the generator, since this would not be allowed in a real experimental arrangement. In the examples given below, the receiver is put at a distance $d$ equal to three times the size of the generator $L_z$.

The magnetic field $\Bdet_0$ in the receiver will amplify the component $\Bdet^{(1)}_x$ of the induced magnetic field, given by the potentials $A^{(1)}_{y,z}$ which are generated by $\textsl{J}_{y,z}$. For the TEM wave resonance generator (a), we therefore only need to compute $\partial_z h_{xx}$ from Eq.(\ref{jeff_TEM}) while for the TM wave resonance generator (b), we need to compute $\partial_{y,z} h_{23}$ from Eq.(\ref{jeff_TM}). 

Finally, in experiment (a) we place the receiver at a distance in the $y$ direction equal to three times the length of the generator  while in experiment (b) at the same distance but in the $z$ direction (see also Figs. \ref{fig_exp_TEMres} and \ref{fig_exp_TMres}). The relative positions of the emitter and the receiver influence the shape of the gravitational wavefront that enters the receiver, hence impacting the volume integral of $||\vec{\Bdet}^{(1)}||$ in Eq.(\ref{dE}). These choices were made for the sake of illustration and do not impact the results presented here.

The numerical resolution of the 3D wave equation Eq.(\ref{inverse_GZ_adim}) is made with Fourier method so that the gradients of $A^{(1)}_{y,z}$ can easily be computed and then averaged to evaluate the energy variation $\Delta E$ from Eq.(\ref{dE}).
\begin{figure}[ht!]
\includegraphics[trim={0cm 0cm 0cm 0cm},clip,scale=0.3]{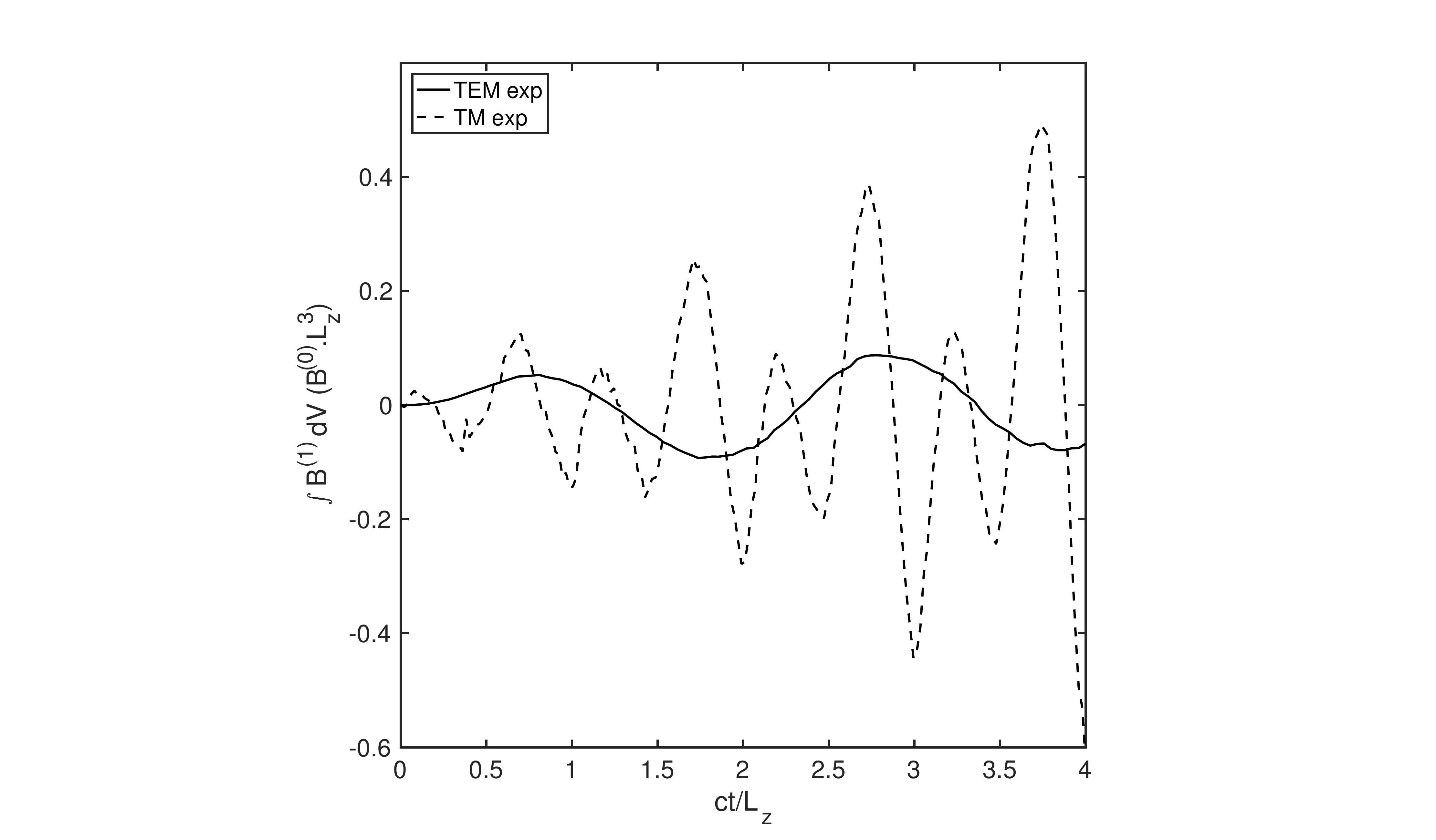}
\caption{Volume integral of $||\vec{\Bdet}^{(1)}||$ in the receiver as a function of time}
\label{fig_deltaE}
\end{figure}

This numerical computation allows us to compute the average of 
$||\vec{\Bdet}^{(1)}||$ over the volume of the receiver:
$$
\left< ||\vec{\Bdet}^{(1)}||\right>=\frac{1}{V_{\rm rec}}\int_{\rm rec}||\vec{\Bdet}^{(1)}||dV
$$
 such that the absolute variation of magnetic energy stored in the receiver is given by
 $$
 \Delta E\approx\frac{\Bdet^{(0)}\left< ||\vec{\Bdet}^{(1)}||\right>V_{\rm rec}}{\mu_0}\cdot
 $$
The numerical results of the time evolution of $\left< ||\vec{\Bdet}^{(1)}||\right>$
for both experiments (a,b) are given in Fig. \ref{fig_deltaE} where one can see that the absolute variation of magnetic energy stored in the receiver undergoes oscillations with increasing amplitude as the energy induced by the passing GWs is accumulated in the receiver. We see that the estimation Eq.(\ref{dEbis}) provides a good order of magnitude of the energy variation in the receiver. These numerical results depend on the relative position of the emitter and the receiver, but the evolution stays qualitatively the same, and in agreement with the analytical results of section \ref{planeGW}.

Let us now point out some technical difficulties that can be expected in translating this ideal concept into a real-life experiment.
Of course, the discussion below is not exhaustive and could not replace a future thorough experimental feasibility study.
The principal difficulty seems to us that any laboratory experiment such as those, involving Einstein field equations, requires a large amount of energy stored into the smallest scale possible, hence reaching a significant compactness,  and will produce tiny physical effects to be measured in a highly energetic environment. For a GW emission-reception experiment, the generators must be the most powerful possible while the detectors must be the most sensitive possible. Emitter and receiver must be put as close as possible, which is limited by the magnet yokes and return paths. There is also the issue of the EM crosstalk, due to the overlap of the EM fields of the emitter and the receiver, which can perturb measurements. However, a GW signal is expected to be detected only when the GW generator is active, i.e. when both the outer magnetic field and the EM standing waves in the cavity are present. Therefore, the EM influence of the generator on the detector can be studied in the absence of GW signal by simply disabling either the outer magnetic field or the EM standing waves. Since the experiment will likely require heavy equipment, one can worry about the influence of
the gravitational potential of these on the GW detection process. This does not seem to us a serious issue because the effect of the static gravitational potential can also be isolated by studying the detection process when the GW generator does not emit. Would the practical realisation of this experiment involve superconducting cavities for the GW generator, then operating this superconducting cavity in a strong outer magnetic field, that should be kept under-critical, might constitute a serious technical challenge. Problems with mechanical vibrations of the resonant cavities and stresses flexing them  during operation have been studied in \cite{grishchuk1} and their conclusion is that these effects can be neglected. In addition, our detecting scheme does not involve any resonant cavity.

\section{Conclusion} 

Controlling gravity, for instance through producing then detecting artificially generated gravitational waves, does not require any new physics nor technology. Indeed, it is in principle achievable within standard general relativity, notably through the Gertsenshtein-Zeldovich effect, and the use of high-field magnets surrounding electromagnetic  waveguides and resonant cavities. 

However, gravitation is an extremely weak interaction compared to electromagnetism and nuclear forces. In fact, generating gravitational fields requires to store large amounts of energy on a short scale.
Indeed, the amplitude $h$ of the GWs generated from EM devices is of order of magnitude of the following dimensionless constant:
\be
h\sim \frac{G E_{\rm EM}}{c^4 L}\label{compactness}
\ee
with $G$ the Newton constant, $c$ the speed of light, $L$ the characteristic size of the generator and 
$E_{\rm EM}$ the total amount of EM energy stored in the GW generator. This dimensionless quantity is analogous to the compactness $s=GM/(c^2R)$ of a compact object of size $R$ filled with inertial mass $M$. 
This argument Eq.(\ref{compactness}) gives a nice interpretation of the Gertsenshtein-Zeldovich number Eq.(\ref{GZ}) in terms of EM compactness. Since our technology presently cannot 
\textit{compactly store energy} as efficiently as nature does in self-gravitating objects, we are only able to produce tiny gravitational fields and detecting them is a challenge for high-precision physics.

The present paper has examined this question of manipulating gravitational fields in the lab through the introduction of three new improved designs of EM generators of GWs.
We have also examined in details the interplay of the EM and GW polarizations in the emission process and the directional propagation of the emitted GWs.
 We have also developed a new method for directional reception of GW using intense magnetic fields and magnetic energy storage. We have proposed three possible applications under the form of experimental concepts for the development of GW physics in the laboratory. First, one could build an experiment aiming at
 detecting indirectly the emission of GWs through the cumulated energy loss in the generator. The detection threshold seems to be reachable since it is of the same order of magnitude as the sensitivity of axion dark matter searches \cite{ADMX}, developed in a similar experimental scheme. Second, we show how a magnetic energy storage system can be used as a directional detector of GW coming from outer space, which could be realised with the promising technology of superconducting magnetic energy storage \cite{smes}. Finally, we have applied these results to the controlled emission and reception of GWs, a gravitational counterpart of the Hertz experiment for electromagnetism already present in \cite{weber,grishchuk2,grishchuk3,rudenko}. We showed that the detection threshold can be reached, although with large and very sensitive experimental set-ups and with expected challenging technical difficulties.
 \\
 \\
Finally, we claim that such gravity control experiments, through the laboratory production of gravitational fields from pure electromagnetic sources, will constitute a unique test of the equivalence principle. Indeed, this involves relativistic sources in the weak gravitational field limit while other tests of the equivalence principle mostly involve (i) non-relativistic sources (inertial masses), (ii) composite systems made of different types of binding energies (chemical, nuclear, electromagnetic, gravitational, etc.) and (iii) permanent gravitational fields due to some inertial mass. Implementing gravity experiments as is envisioned here will offer a unique opportunity to test the equivalence principle in the weak field regime but for relativistic sources made of pure electromagnetic energies producing oscillating deformations of space-time.\\
\\
Harnessing gravity, the last indomitable fundamental interaction, constitutes a true experimental challenge, but will undoubtedly lead to
 rewarding scientific breakthroughs and new technologies.
 \\
 \\
\noindent \textit{Acknowledgments}
This research used resources of the "Plateforme Technologique de Calcul Intensif (PTCI)"
(http://www.ptci.unamur.be) located at the University of Namur, Belgium, which is supported 
by the F.R.S.-FNRS under the convention No. 2.5020.11. 
 
\end{document}